\documentclass[useAMS]{mn2e}
\usepackage[dvips]{graphicx}
\usepackage{amssymb,amsmath}
\usepackage{color}
\usepackage{url}

\title[Angular momentum flips and galaxy formation]{Stochastic angular momentum slews and flips and their effect on discs in galaxy formation models}
\author[N.D. Padilla et al.]{\parbox[t]{\textwidth}{\vspace{-1cm}
Nelson D. Padilla$^{1,2}$\thanks{E-mail:npadilla@astro.puc.cl},
Salvador Salazar-Albornoz$^{3,4,1}$, Sergio Contreras$^1$, Sof\'\i a A. Cora$^{5,6,7}$, Andr\'es N. Ruiz$^{7,8,9}$}\\
$^{1}$Instituto de Astrof\'\i sica, Pontificia Universidad Cat\'olica de Chile, Santiago, Chile\\
$^{2}$Centro de Astro-Ingenier\'ia, Pontificia Universidad Cat\'olica de Chile, Santiago, Chile\\
$^{3}$Universit\"ats-Sternwarte M\"unchen, Scheinerstrasse 1, 81679 Munich, Germany\\
$^{4}$Max-Planck-Institut f\"ur extraterrestrische Physik, Giessenbachstrasse 1, 85748 Garching, Germany\\
$^{5}$Instituto de Astrof\'isica de La Plata (CCT La Plata, CONICET, UNLP), Paseo del Bosque s/n, B1900FWA, La Plata, Argentina.\\
$^{6}$Facultad de Ciencias Astron\'omicas y Geof\'{\i}sicas, Universidad Nacional de La Plata, Paseo del Bosque s/n, B1900FWA, La Plata, Argentina.\\
$^{7}$Consejo Nacional de Investigaciones Cient\'{\i}ficas y T\'ecnicas,
Rivadavia 1917, Buenos Aires, Argentina\\
$^{8}$Instituto de Astronom\'{\i}a Te\'orica y Experimental (CCT C\'ordoba, CONICET, UNC), Laprida 854, C\'ordoba, X5000BGR, Argentina\\
$^{9}$Observatorio Astron\'omico de C\'ordoba, Universidad Nacional de C\'ordoba, Laprida 854, C\'ordoba, X5000BGR, Argentina
}

\begin{document}

\date{Accepted --. Received --; in original form --}

\pagerange{\pageref{firstpage}--\pageref{lastpage}} \pubyear{2013} %change!!!

\maketitle

\label{firstpage}

\begin{abstract}
{ The angular momentum of galactic discs in semi-analytic models of galaxy formation is usually updated in time as material is accreted to the disc by adopting a constant dimensionless spin parameter and little attention is paid to the effects of accretion with misaligned angular momenta.  These effects are the subject of this paper, where we adopt a Monte-Carlo simulation for the changes in the direction of the angular momentum of a galaxy disc as it accretes matter based on accurate measurements from dark-matter haloes in the Millennium II simulation. In our semi-analytic model implementation, the flips seen the dark matter haloes are assumed to be the same for the cold baryons; however, we also assume that in the latter the flip also entails a difficulty for the disc to increase its angular momentum which causes the disc to become smaller relative to a no-flip case.  This makes star formation to occur faster, specially in low mass galaxies at all redshifts allowing galaxies to reach higher stellar masses faster.  We adopt a new condition for the triggering of starbursts during mergers.  As these produce the largest flips it is natural to adopt the disc instability criterion to evaluate the triggering of bursts in mergers instead of one based on mass ratios as in the original model.  The new implementation reduces the average lifetimes of discs by a factor of $\sim 2$, while still allowing old ages for the present-day discs of large spiral galaxies.  It also provides a faster decline of star formation in massive galaxies and a better fit to the bright end of the luminosity function at $z=0$.}
\end{abstract}


\section{Introduction}
\label{sec:intro}

The formation and evolution of galactic discs in a universe dominated by dark matter poses
great challenges that still need to be solved.  For instance, only recently grand design discs such as that
of the Milky Way have been produced in hydrodynamical simulations with a better
treatment of multiple gas phases and feedback (Scannapieco et al., 2006) and with the hybrid 
lagrangian cell code AREPO (Springel, 2010).  On the other hand, it is difficult to produce large enough
samples of galaxies with these codes to make statistical comparisons with galaxy samples
extracted from large surveys such as the Sloan Digital Survey (York et al., 2000) due to
the large computational resources required for this
in comparison to simulations that only follow dark matter (see for instance Bower, Benson \& Crain, 2012).

In recent years, these problems have been partially solved by different approaches such as
(i) using resimulations of chosen haloes in large
dark matter only simulations to much higher resolutions, as is the case of the Aquarius simulations (Springel et al., 2008).
These haloes resemble our own Milky Way$'$s (MW), and simulations of galaxies within these are expected
to reproduce the large disc of the MW at least in some cases.  These simulations have been extensively analyzed and
different hydrodynamical codes have been run using the Aquarius simulations (the Aquila project, Scannapieco et al., 2012), 
and in some cases their results are promising showing grand-design spiral galaxies in a fraction of the 
haloes (Marinacci, Pakmor \& Springel, 2013).
(ii) Resimulating larger portions of the Millennium Simulation (Springel et al. 2005) corresponding to different
environments, of low, average and high matter density; this is the case of the GIMIC project (Galaxies-Intergalactic
medium Interaction Calculation, Crain et al., 2009).  However, it turned out that
the resulting stellar mass function of the GIMIC
galaxy population is not compatible with the observed one (Bower, Benson \& Crain, 2012).
The limitation of this approach is evident since due to
the high computational demand of this project,
 it was not possible to re-run it many times with
changing parameters until the resulting galaxy population matched the observed one.
(iii) Currently, the EAGLE project (Evolution of Galaxies and their Environment, Bower, Benson \& Crain, 2012) 
seems the most promising way to obtain fully hydrodynamical simulations
of a galaxy population with the right stellar mass function, that is, with a reasonable galaxy population
in which each galaxy has evolved embedded in a dark matter halo, with all the associated physical
effects, including those involved in
the formation and evolution of galactic discs.

{ Still, even though cosmological volumes are being simulated using reasonable hydrodynamics, in some
cases there are still issues regarding the resulting efficiency of early star formation, which can be
higher than observed (e.g.  Powell et al., 2011, Brook et al., 2011, Nagamine 2010, and references therein).  
The proposed solution to this appears to come from adopting higher resolution and introducing 
additional sources of feedback (e.g. Hopkins et al., 2011, Tasker 2011, Kannan et al., 2013).
}

Another solution to the problem of making large galaxy samples from simulations comes from
semi-analytic models of galaxy formation.  These models 
are necessarily extremely simplified versions of the hydro simulations, since  
a small set of simple equations describes
the evolution of, for instance, an entire gas phase in a galaxy (Kauffmann et al., 1999,
Cole et al., 2000, see also the reviews by Baugh, 2006, Benson, 2010, Silk \& Mammon, 2012);
this is, however, the reason why it is feasible to generate large galaxy populations with these
models.  
In particular, the process of star formation is generally
modeled according to the observed
proportionality between star formation rate and projected
gas density (e.g. Kennicut, 1998).
{  
The proportionality constant involves an efficiency parameter, which is fixed so
as to fit the observed z = 0
galaxy luminosity function, among other properties.
Notice that the early over-efficiency of star formation affecting some hydrodynamical 
simulations is therefore not present in semi-analytic
models, since they directly fix the efficiency of star formation by reproducing the observed
total mass in stars at $z=0$.
}

The ingredients
involved in the star formation process in semi-analytics 
are the disc dynamical time and the mass of cold gas contained in the disc
(e.g. Cole et al., 2000,
Croton et al., 2006, Lagos, Cora \& Padilla, 2008, among others).  Of the two, the dynamical time is the most
delicate in the sense that it depends strongly on the size of the disc, and this is a challenging quantity as
it involves many complicated evolutionary processes.  
For instance, we need to know the angular momentum of the disc
and which is the influence of the bulge in its final size.
Practical answers to these questions
were given by Mo, Mao \& White (1998, MMW), by studying the behaviour of discs in numerical simulations.  For
relaxed haloes, they were able
to provide formulae for the typical disc size that depend on the dark matter halo mass, its specific angular
momentum, $\lambda$, the fraction of mass residing in the disc, and the fraction of mass in the bulge.

However, haloes acquire mass continuously, 
either in a smooth way or via mergers, and in both cases, the angular 
momentum of the halo suffers slews and flips, 
some of them small but others as high as $90$ degrees or even more.
Bett \& Frenk (2012)
studied the frequency of flips in dark matter only simulations.
They showed that flips are more likely in
mergers, but that they still occur when there is smooth accretion.  
If the hot gas is relaxed within
the dark matter halo, as it cools it is highly likely that the disc sitting in the
center will have a different angular momentum. %the disc spin will point in a different
Analysing the GIMIC galaxies, Sales et al. (2012) showed that this effect can 
destroy discs completely, 
making their life more episodic than is usually obtained in semi-analytic models, where
unless a merger 
or a disc instability
takes place, the disc keeps growing (on average).
In the same line, Aumer et al. (2013) find that the last misaligned accretion influences the disc fraction
of a galaxy.
Other studies of disc galaxies show the complex physics of discs.  Saha \& Naab (2013) find that the alignment
of the angular momentum direction of the disc and halo produce stronger bars, which in principle could suggest
that not even aligned discs are guaranteed to survive long.  In a more cautious note, Bird et al. (2013)
analysed the Eris ensemble of resimulated haloes (Guedes et al., 2011), and point
out that initial conditions have an important influence on the final state of the disc of a galaxy.

The aim of this paper is to evaluate the impact on the properties of
galactic discs of the angular momentum slews and flips suffered by dark matter halos
using the semi-analytic model of galaxy formation {\small SAG}
(Cora, 2006;
Lagos, Cora \& Padilla, 2008; Lagos, Padilla \& Cora, 2009a;
Tecce et al. 2010).
Although the version of the model described in Tecce et al. (2010),
to which we will refer as the base model, makes 
full use of the MMW formulae to determine disc sizes,
it does not take into account the change of the angular momentum of the discs as they accrete matter. 
In this paper we will improve upon this recipe by taking this effect into account, to some degree, 
developing a new version of this model that we will refer to as the flip model.
For this purpose,  
we construct a statistics of angular momentum slews using the Millennium II
simulation (Boylan-Kolchin et al., 2009) and, with the aid of a Monte-Carlo simulation,
apply slews in the discs of the semi-analytic galaxies.
The resulting properties of galactic discs impact directly the frequency
of disc instability events suffered by a galaxy,
in which the disc is transferred to the galactic bulge with all
the cold gas available being consumed in a starburst.
The chance of bursts increases with the amplitude of the flip of the angular momentum
of the discs, which makes us
implement one additional modification to our model 
regarding the triggering of starbursts during
galaxy mergers. 
Instead of using mass ratios of the merging galaxies
to classify them as major mergers with starburst or minor mergers with and
without starbursts, as is  
considered in the base model, we now apply the disc instability criterion 
to the remnant galaxy.  

One of the possible consequences of 
introducing angular momentum flips in galaxy discs in the model
is that it could affect the rate
at which star formation occurs at high redshift.  Since a few years back there has been some tension
between the observed abundance of massive galaxies at high redshift and those predicted by semi-analytic
models (e.g. de Lucia et al., 2006). Several observational studies pointed at larger populations
of high stellar mass galaxies than predicted (e.g. P\'erez-Gonz\'alez et al., 2008, Ferreras et al., 2009),
but with the advent of mid-infrared imaging of high redshift galaxies, this situation was somewhat
alleviated, as shown by Marchesini et al. (2010).  They point out that if dusty templates are included
in the analysis of photometric samples of galaxies with mid-IR coverage, then the tension with the 
models is reduced.  With an increased star formation activity at high redshifts, our new model could help to reduce
even further this controversy.

This paper is organised as follows.  In Section 
\ref{sec:flips} we present the statistics on angular momentum slews extracted from the Millennium II
simulation (Boylan-Kolchin et al., 2009), 
and we briefly describe the aspects of the semi-analytic model
relevant to the present study 
and the improvements introduced 
regarding the slews of dark matter angular momentum;
Section \ref{sec:di} shows
the effects of these 
specific changes
on the typical lifetimes of discs in galaxies.
Section \ref{sec:mergers} shows how with this new model we can follow
bursts in mergers using only the disc instability criterion instead of mass ratios as in the base model.
In Section \ref{sec:results} we explore the consequences of this new treatment on the
general properties of the model galaxy population, and Section \ref{sec:conclusions} summarises our results.

\section{Angular momentum slews and flips in a semi-analytic model}
\label{sec:flips}

\begin{figure}
  \begin{center}
    \includegraphics[width=.44\textwidth]{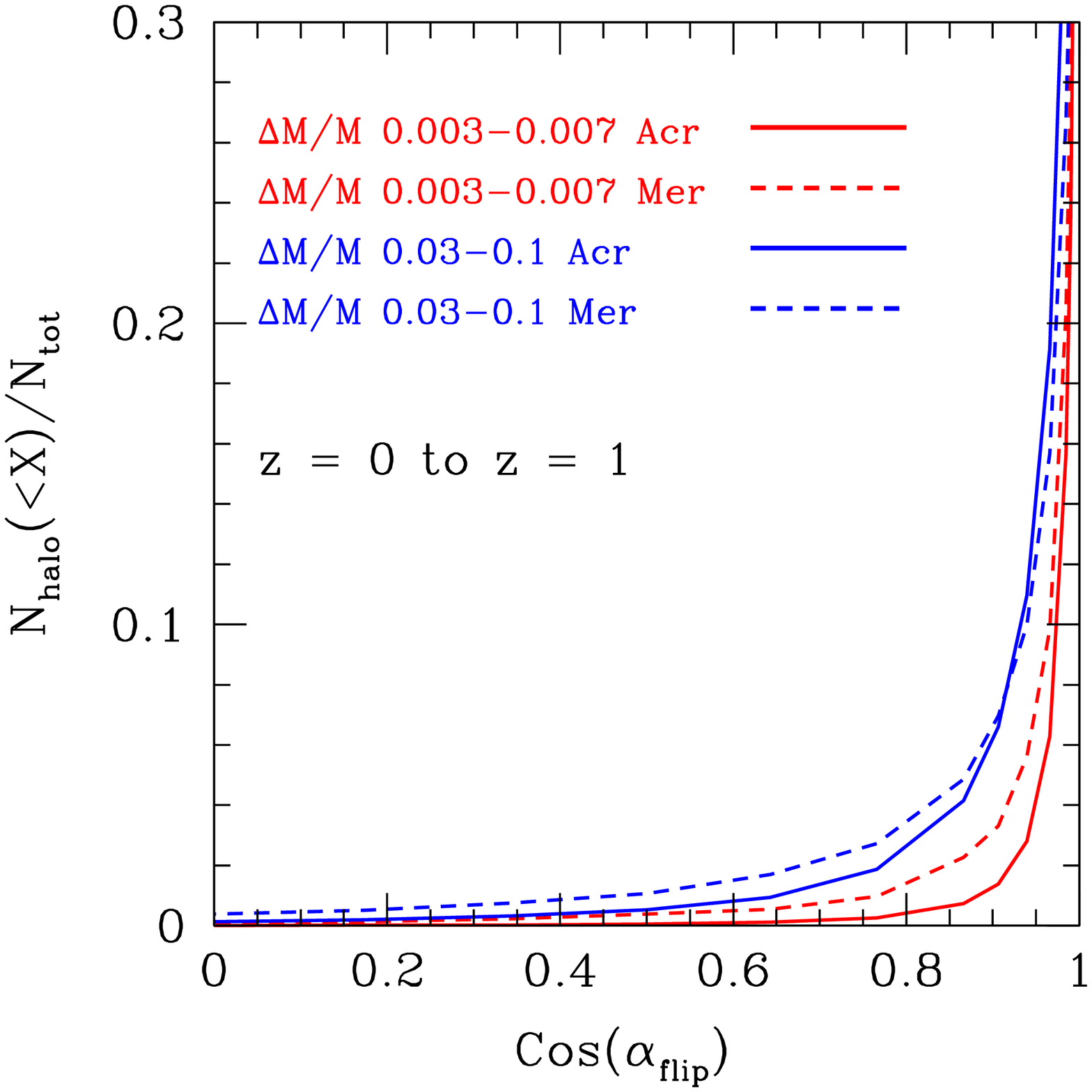}
    \includegraphics[width=.44\textwidth]{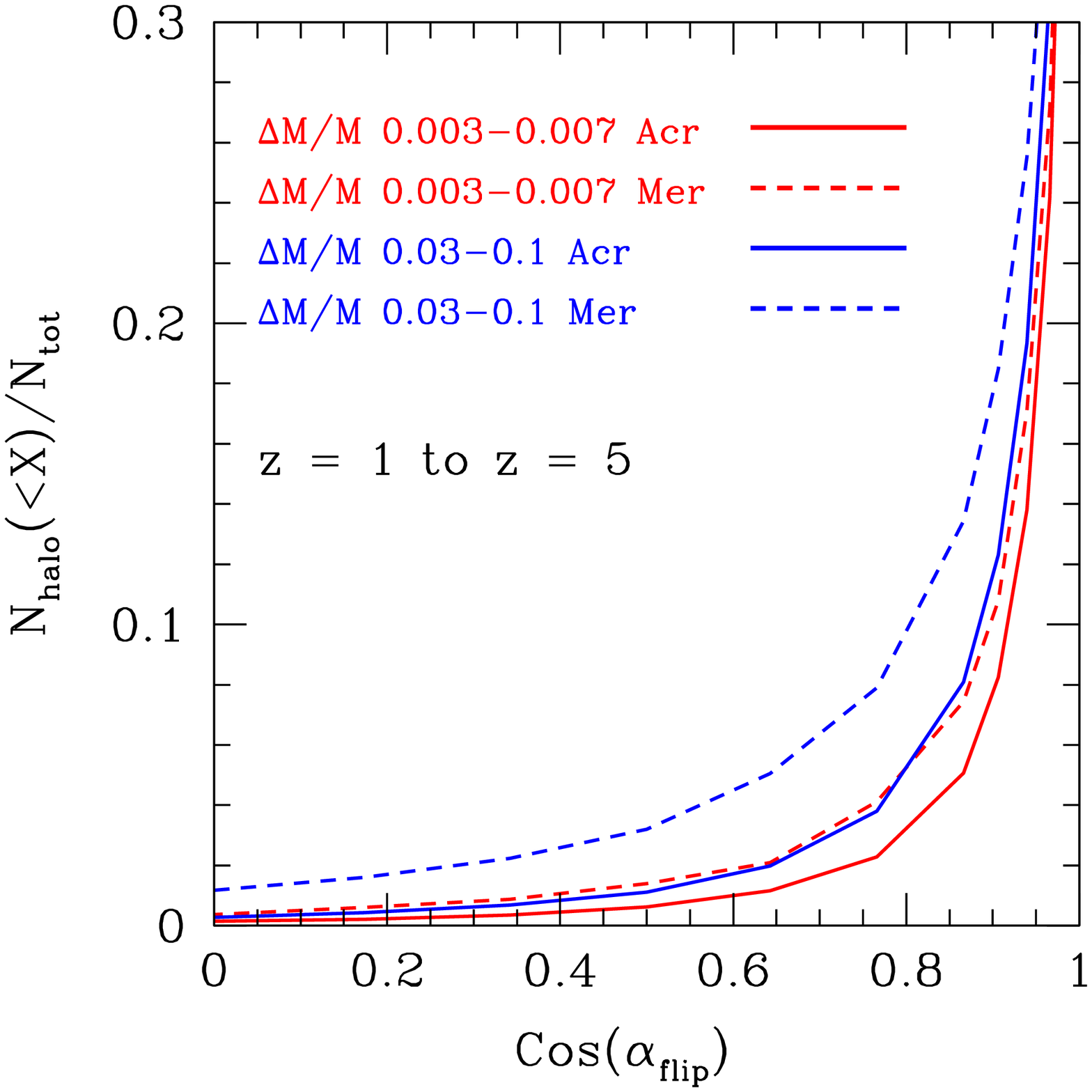}
    \caption{
Cumulative distributions of the cosine of $\alpha_{\rm flip}$, the angle between the angular momenta of a dark matter substructure
in the Millennium II simulation 
before and after
accreting matter; these angular momentum changes are commonly
referred to as slews or flips.  
Top and bottom panels correspond to accretion events
between $0<z<1$ and $1<z<5$, respectively.  
Dashed lines show the distribution of
flips in merger events, whereas solid lines are for accretion of individual 
dark matter particles. The
different colors correspond to two different { representative} ranges in relative mass accretion, corresponding to
$0.003<\Delta M/M<0.007$ and
$0.03<\Delta M/M<0.1$, as indicated in the figure key.
     \label{fig:angmom}}
  \end{center}
\end{figure}

We use the {\small SAG} (Semi-Analytic Galaxies) semi-analytic model of galaxy formation which is based 
on the 
one described by
Springel et al. (2001), with improvements on the chemical element production by
Cora (2006), the implementation of Active Galactic Nucleus (AGN) feedback in Lagos, Cora \& Padilla (2008), and 
the inclusion of MMW disc sizes by Tecce et al. (2010).
{\small SAG} is run on subhalo merger trees extracted from a dark-matter only N-body simulation.
This simulation is based on the standard $\Lambda$CDM scenario,
characterized by
the cosmological parameters $\Omega_{\rm m}=0.28$,
$\Omega_{\rm b}=0.046$, $\Omega_{\Lambda}=0.72$, $h=0.7$, $n=0.96$,
$\sigma_{8}=0.82$, according
to the WMAP7 cosmology (Jarosik et al., 2011). The simulation was
run using GADGET-2 (Springel 2005)
using $640^3$ particles in a cubic box of comoving sidelength
$L=150\,h^{-1}{\rm Mpc}$.

In this paper, we 
introduce angular momentum slews and flips that will directly affect the galaxy discs.  One simple
way to do this is to follow the angular momentum vector of the 
dark matter
halo which changes with time, and assume that
the hot gas always carries it, even when cooling down to top up the cold gas supply of the disc.  The incoming cold
gas brings angular momentum parallel to that of the halo, but if the halo changed its spin with respect to the
time when the disc was formed, the cooling gas will have a different angular momentum than the disc.  Using mass weighted
averages 
of the angular momenta of the incoming cold gas 
and the galaxy disc,
Lagos, Padilla \& Cora (2009a) followed the changes in the angular momentum of the galaxy disc.  They went even further
and extended this to the spin of the black hole (see also Lagos et al., 2011, and Guo et al., 2011).

However, the direction of the angular momentum of haloes is difficult to measure accurately; for haloes with less than $1000$
particles the direction is subject to large contributions from numerical noise (e.g. Bett \& Frenk, 2012).  Therefore,
the changes in the direction of the angular momentum of haloes in this case is larger than it would be if the haloes
were followed with higher resolution.  In our case, therefore, we either need to smooth this effect to some degree
(as was done in Lagos, Padilla \& Cora, 2009a), or use a statistical measurement of the typical angular momentum
flips of haloes to apply it to a semi-analytic model.

We opt for the latter, for which the first step is to obtain a full statistical distribution of flips from
a numerical simulation with enough resolution so that the full dynamical range of haloes populated by {\small SAG} is
covered by our statistics.  The Millennium II simulation (Boylan-Kolchin et al., 2009) provides us with
subhaloes of $M>10^{10}$h$^{-1}M_{\odot}$ with more than $1000$ particles each, fulfilling this requirement. 
\footnote{The Millennium II simulation data were obtained using the German Astrophysical Virtual Observatory
GAVO, at http://gavo.mpa-garching.mpg.de/MyMillennium/}

We will use subhaloes instead
of haloes from the start, to avoid the influence in the measured angular momentum from other substructures.  
This way
satellites { will slew according to the angular momentum of their host substructures and not of their host} halo as a whole. 
Furthermore, central galaxies are this way unaffected by the changes in the angular momentum of their 
satellites, which should not contribute to the central galaxy$'$s 
angular momentum until the event of a merger.

Figure \ref{fig:angmom} shows the probability distributions for the cosine of the angle between the initial
and final angular momentum
($\alpha_{\rm flip}$)
 of subhaloes that suffer changes in their mass by the relative amount indicated
in the key { (these relative mass ranges are chosen as examples, the actual statistics cover a wide
range of mass accretion ranging up to $\Delta M/M>0.5$)}.  
These angular momentum changes are commonly
referred to as slews or flips.  
Different line types correspond to whether the increase in mass was by a merger (dashed lines)
or smooth accretion (solid lines).  
The statistics for smooth accretion are obtained analysing subhaloes that have not suffered
mergers in consecutive snapshots of
the simulation.  To construct this for the case of mergers, we take only subhaloes that are the result
of at least one merger since the previous snapshot and measure their flips.  { In the case when the
merged subhalo mass is larger than the sum of its progenitor masses, we assume that the excess mass came from
smooth accretion.  In this case we 
lower the amplitude of the flip from mergers using the statistics for smooth accretion
which are measured first.  Only after this correction is made we use it for the statistics of mergers.}
The upper solid (dashed) line indicates that a higher relative mass accretion
(merger) increases the chances of a given angular momentum flip.  The upper panel corresponds to low redshifts, $0<z<1$,
whereas the lower panel to $1<z<5$, and as can be seen, the chance of angular momentum flips is higher
at higher redshifts.

\subsection{Episodic discs}

\begin{figure}
  \begin{center}
    \includegraphics[width=.44\textwidth]{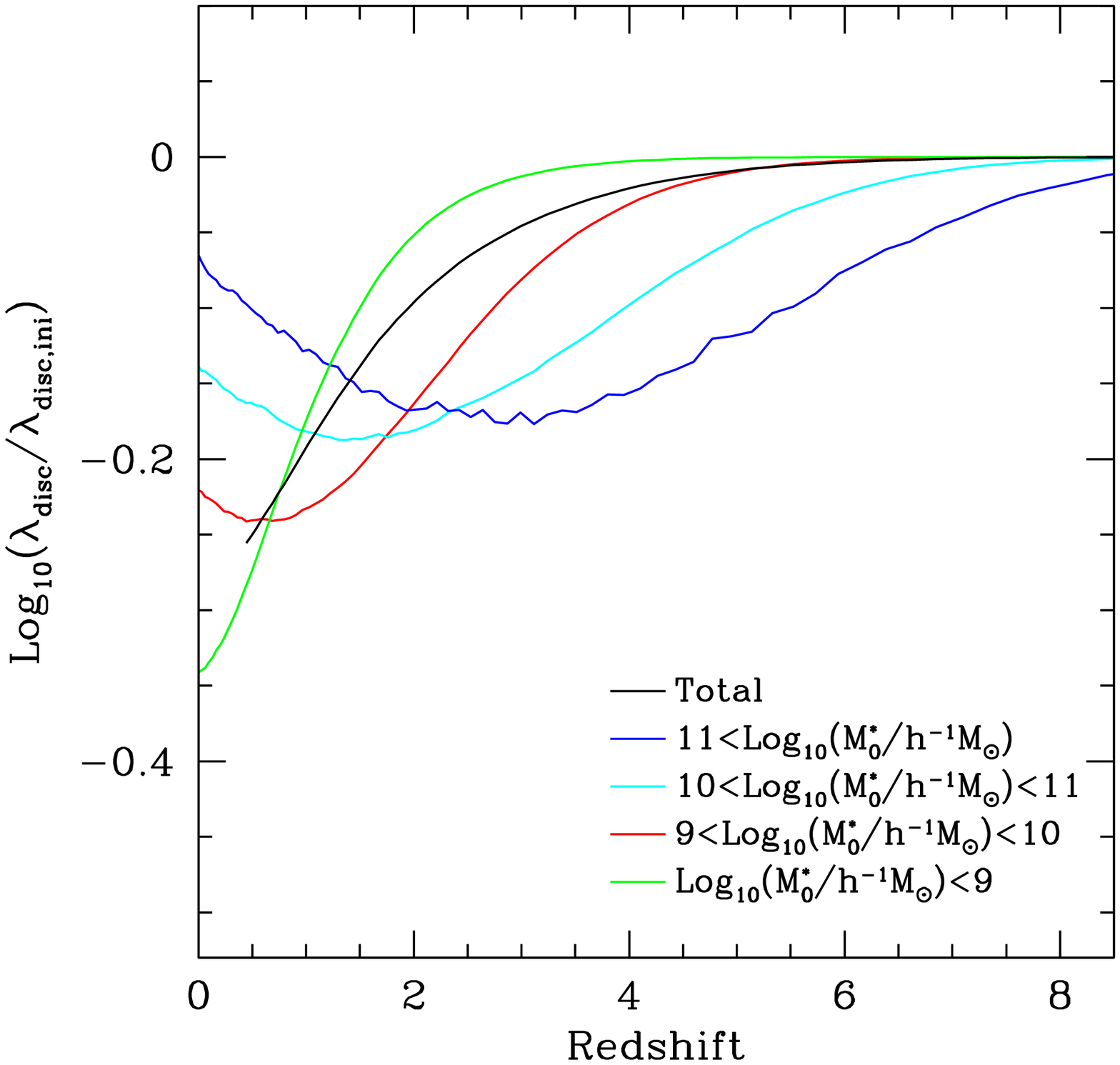}
    \includegraphics[width=.44\textwidth]{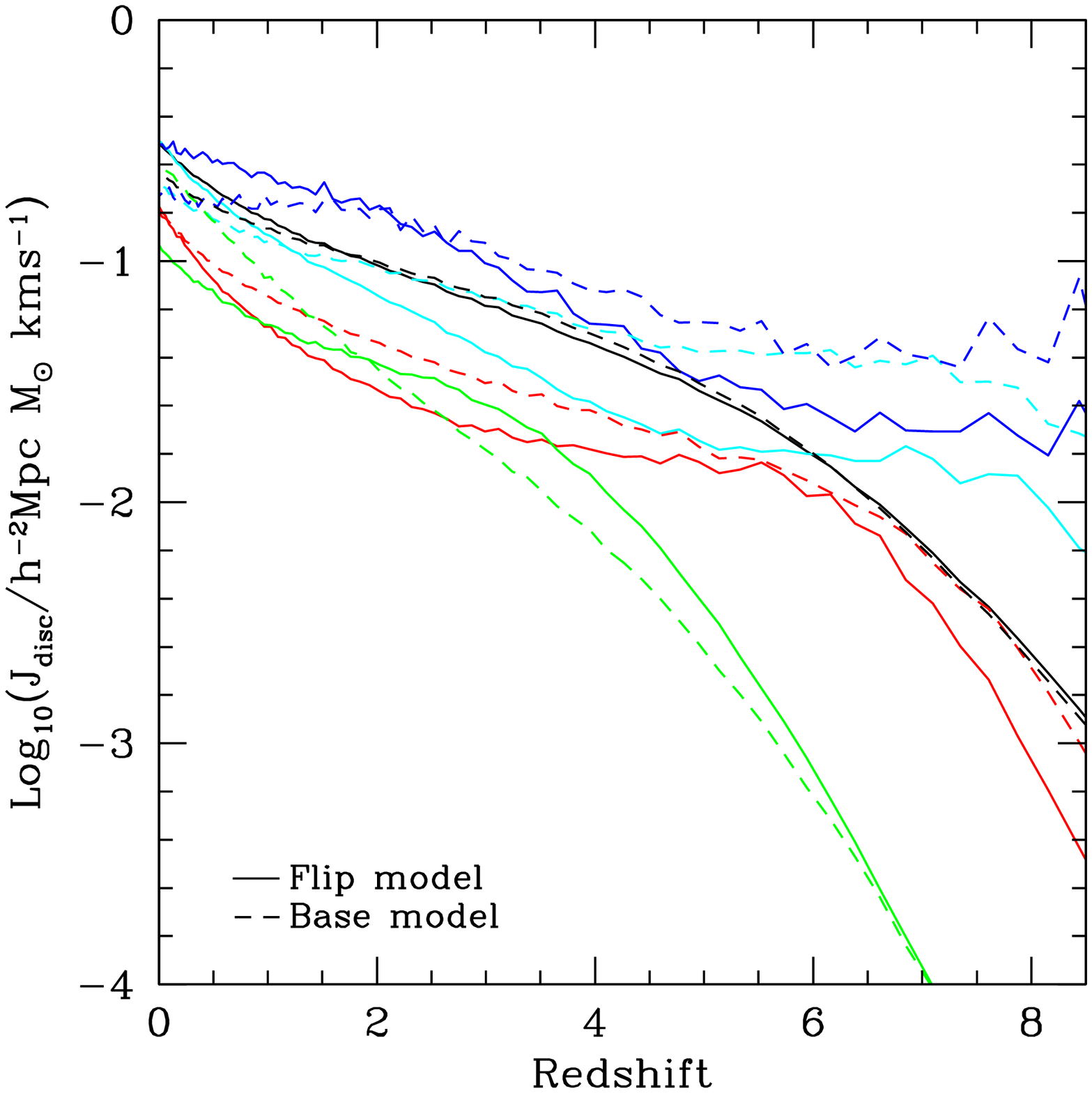}
    \caption{Top: average $\eta_{disc}$ values as a function of redshift for galaxies with present-day stellar masses
in four different ranges, as shown by the different colours indicated in the figure key. Bottom:  
average angular momentum of galaxy discs as a function of redshift, for the same 
present-day stellar mass ranges as the top panel (different colours), for the base and flip model shown
as dashed and solid lines, respectively.
}
   \label{fig:etadisk}
  \end{center}
\end{figure}

Our application of the probability of slews and flips as a way to follow the evolution of the angular momentum
of semi-analytic galaxy discs is simple.  In each snapshot of the simulation, we calculate
the fractional increase in the galaxy disc mass.  We then assign a disc angular momentum flip following
the statistical distributions measured from the Millennium II subhaloes, assuming that the same slews or
flips are suffered by the baryons in their centres.  

However,
the most important assumption is that { when the disc suffers a flip
that amounts to an angle 
$\alpha_{\rm flip}$, between the ``old" and ``new" angular momenta, this}
lowers the specific angular momentum of 
the disc by the following amount,
\begin{equation}
\lambda_{\rm disc}=\lambda_{\rm disc,old} \cos(\alpha_{\rm flip}).
\end{equation}
{ The angle $\alpha_{\rm flip}$ accumulates successive slews suffered by an individual disc 
in a statistical way.  As a result, the specific angular momentum of the disc is generally lower than
that of the halo at the time the disc formed, and continues to decrease its value as 
more accretion takes place.  
This is qualitatively consistent with recent findings that point to lower 
specific angular momentum for galaxies in comparison
to that of dark matter haloes (e.g. Dutton \& van den Bosch, 2012).

Hydrodynamical simulations and semi-analytic models suffer from opposite problems.
It has been pointed out that there is a lack of enough angular momentum to form discs in 
hydrodynamical simulations (e.g. Mo, Mao \& White, 1998, Bullock et al., 2001).  Recently,
Robertson et al. (2006) showed that this 
can be overcome in cosmological scenarios by taking into account the effect
of early merging and appropriate feedback mechanisms 
(see also Maller \& Dekel, 2002, D'Onghia \& Burkert, 2004, Vitvitska et al., 2002).  
On the other hand,
these problems are the opposite in semi-analytic models, where it is usually assumed that
the specific angular momentum of the disc equals (e.g. Tecce et al., 2010) 
or is even higher than that of the dark-matter halo (Berry et al. 2013).
This is done because the prescriptions used in these models 
are not able to follow any of the details of angular momentum and feedback energy transfer. 
By assuming these equalities, semi-analytic models adopt
constant specific angular momenta for discs.  However, several works analysing
hydrodynamical simulations (Sales et al., 2012, Saha \& Naab, 2013, Aumer et al., 2013) find that 
the infall of misaligned material tends to disfavour the survival of the disc 
whose destruction could be the result of a change in the properties of the disc such as its spin.

Our proposed loss of specific angular momentum of discs is designed to take this effect into account to
some degree, as it entails an inhibition of growth of the disc angular momentum due to slews and 
flips.} % as long as the cumulative effect of the flips is not too extreme.
The initial specific angular momentum of a disc is exactly that of its host halo as it forms, $\lambda_{\rm disc,ini}$.  Afterward, successive accretion events make 
the specific angular momentum, $\lambda_{\rm disc}$,  smaller with respect to its initial value (without necessarily making the disc actually
lose angular momentum).  We define
\begin{equation}
\eta_{\rm disc}=\lambda_{\rm disc}/\lambda_{\rm disc,ini}.
\end{equation}
The value of $\eta_{\rm disc}$ diminishes
with each accretion event,
but in the case of a merger induced burst of star formation or disc instability, the disc disappears and, therefore,
$\eta_{\rm disc}$ is reset to $1$.  Only after one halo dynamical time has passed the value of $\eta_{\rm disc}$ is allowed
to vary again.  We remark that in this implementation we do not follow the changes in amplitude of $\lambda$
suffered by the host dark matter structure; we only use its value as a new disc forms as in Cora (2006) and 
Lagos, Cora \& Padilla (2008).  
As Guo et al. (2011) pointed out, if the disc
follows the changes of the specific angular momentum of the dark matter its properties will
suffer sudden jumps due to the numerical errors in determining this quantity using few dark matter particles.
We also point out that on average the galaxy population will show a smoothly raising amplitude of
angular momentum of discs, but the effect of the slews and flips on the discs specific angular momentum will change its rate of growth.

{ Notice that even though the amplitude of the dimensionless spin parameter of a disc diminishes, the
disc itself slews the direction around which it is spinning by an angle given by the flips
distribution measured for the dark matter.  This allows us to reproduce, qualitatively,
previous well-known results where discs show changes (sometimes important ones) in their direction
of spin as a result of mass accretion (e.g. Shen \& Sellwood, 2006).  Also, as is shown in Figure
\ref{fig:angmom}, the distribution of the cosine of angle of flips/slews indicates that an accretion
of equal relative mass produces small slews more frequently than actual angular momentum flips.
We also note that our formalism does not directly take into account
the changes of angular momentum of discs due to external torques which can produce tell tale signs
in discs such as warps (e.g. Garcia-Ruiz, Kuijken \& Dubinski, 2002, Dubinski \& Chakrabarty, 2009),
but these processes do not often entail a loss of angular momentum and therefore, to first order, do
not need to be included in our modeling.
Another point to note is that we treat the gas disc as a single object and do not take into account its possible
interaction with the stellar disc, or the existence of counter-rotating stars, or stellar warps and other
features in the distribution of stars (Algorry et al., 2013, Bois et al., 2011, Naab et al., 2013).  Our modeling
should include this mechanism for angular momentum transfer between gas and stars, but its
effect should be smaller than that coming from the much more frequent 
accretion events (e.g. Algorry et al. 2013), 
and we therefore defer its treatment to future work.}

The main motivation behind our implementation comes from the results from hydrodynamical simulations 
that show that the presence of a disc in a galaxy is favoured by a lack of infall 
of material coming with a misaligned angular momentum (e.g. Sales et al., 2012, Saha \& Naab, 
2013, Aumer et al., 2013).  

Figure \ref{fig:etadisk} shows the change in 
the value of $\eta_{\rm disc}$ 
as a function of redshift
for galaxies of different present-day stellar masses, as indicated in the key.  As can be seen, more massive galaxies today are
tending to $\eta_{\rm disc}=1$ values, since their evolution consists of a decreasing number of important accretion events.  However,
at high redshifts, when their masses were still low, their $\eta_{\rm disc}$ 
reached values as low as $0.65$.
Notice that less massive galaxies (present-day) climb out of this minimum value 
of $\eta_{\rm disc}$ at lower redshifts, once they
achieve a certain pivot mass. As a consequence of this, 
the galaxies that today have the lowest $\eta_{\rm disc}$ values are
the least massive ones.

The lower panel of the figure shows the evolution with redshift of the average angular momentum of the disc for the
base and flip model (dashed and solid lines), for the same ranges of present-day stellar masses as in the top panel.
As can be seen, the dips in $\eta_{\rm disc}$ that mark the epoch of large flips 
correspond to the time when the angular momenta slow their rate of growth, as is the case
for the highest present-day stellar masses probed down to $z=6$;  Afterward
$\eta_{\rm disc}$ reaches its minimum value and rapidly increases back to 
unity and the angular momentum grows again.  As can be seen, the angular momentum
of discs in the base model is usually higher than that in the flip model once the galaxies enter their large flips eras. { Stewart et al. (2013) use hydrodynamic simulations to study the evolution of
the angular momentum of the gas in individual disc galaxies; in their Figure 2, it can be seen that
there are several instances where the angular momentum of the gas grows more slowly than that of the dark
matter, and this is the effect that we are reproducing with our simple implementation for the semi-analytic
model.  Regarding observational results, it is interesting to note that at $z=0$ only the flip
model manages to produce  galaxies with higher angular momenta for higher
stellar masses, in qualitative agreement
with the results of Romanowski \& Fall (2012).}

Our assumptions have a number of consequences for the two different cases corresponding to (i) accretion of gas cooling from
the hot phase,
and (ii) mergers.

\begin{figure}
  \begin{center}
    \includegraphics[width=.44\textwidth]{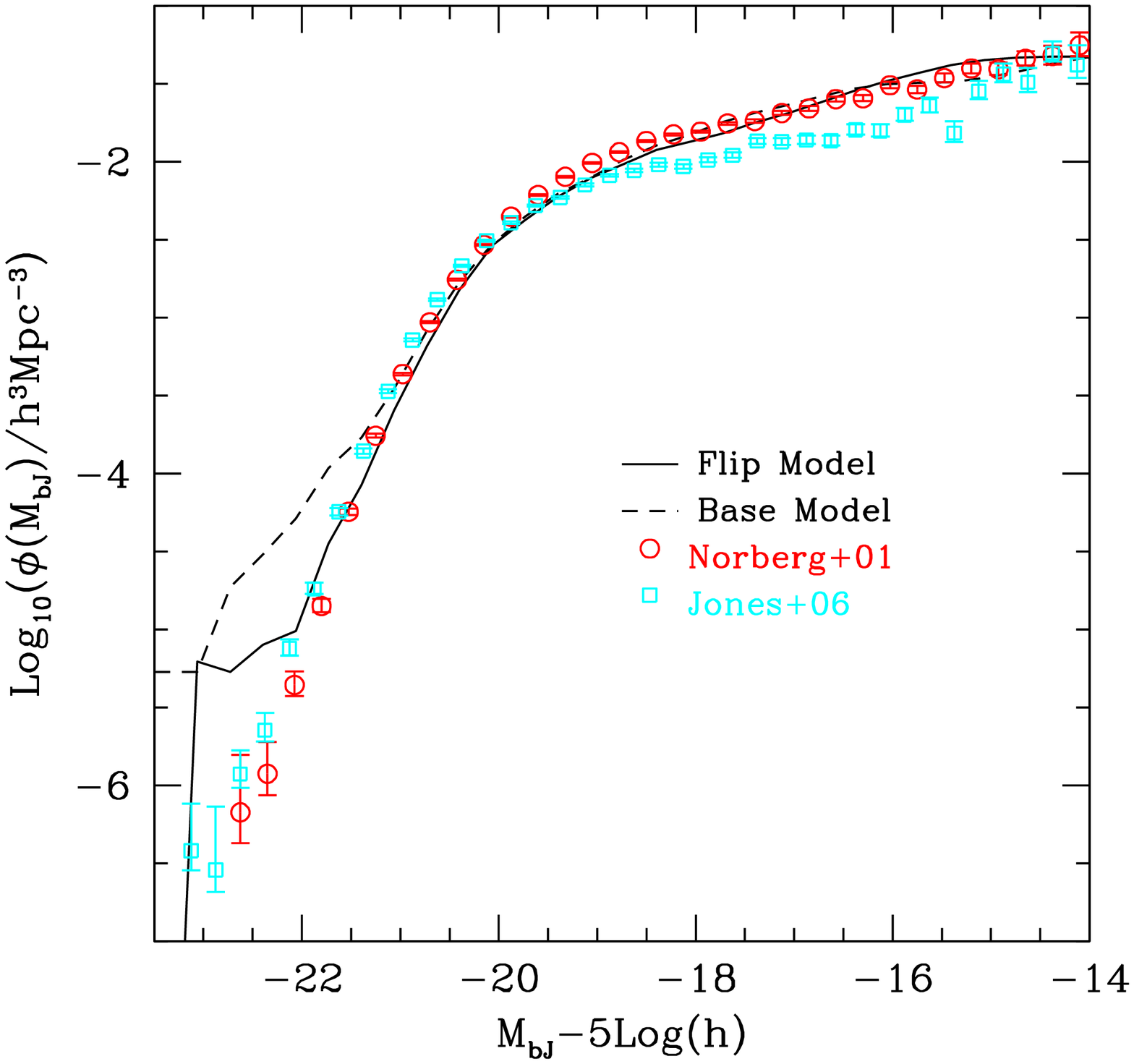}
    \includegraphics[width=.44\textwidth]{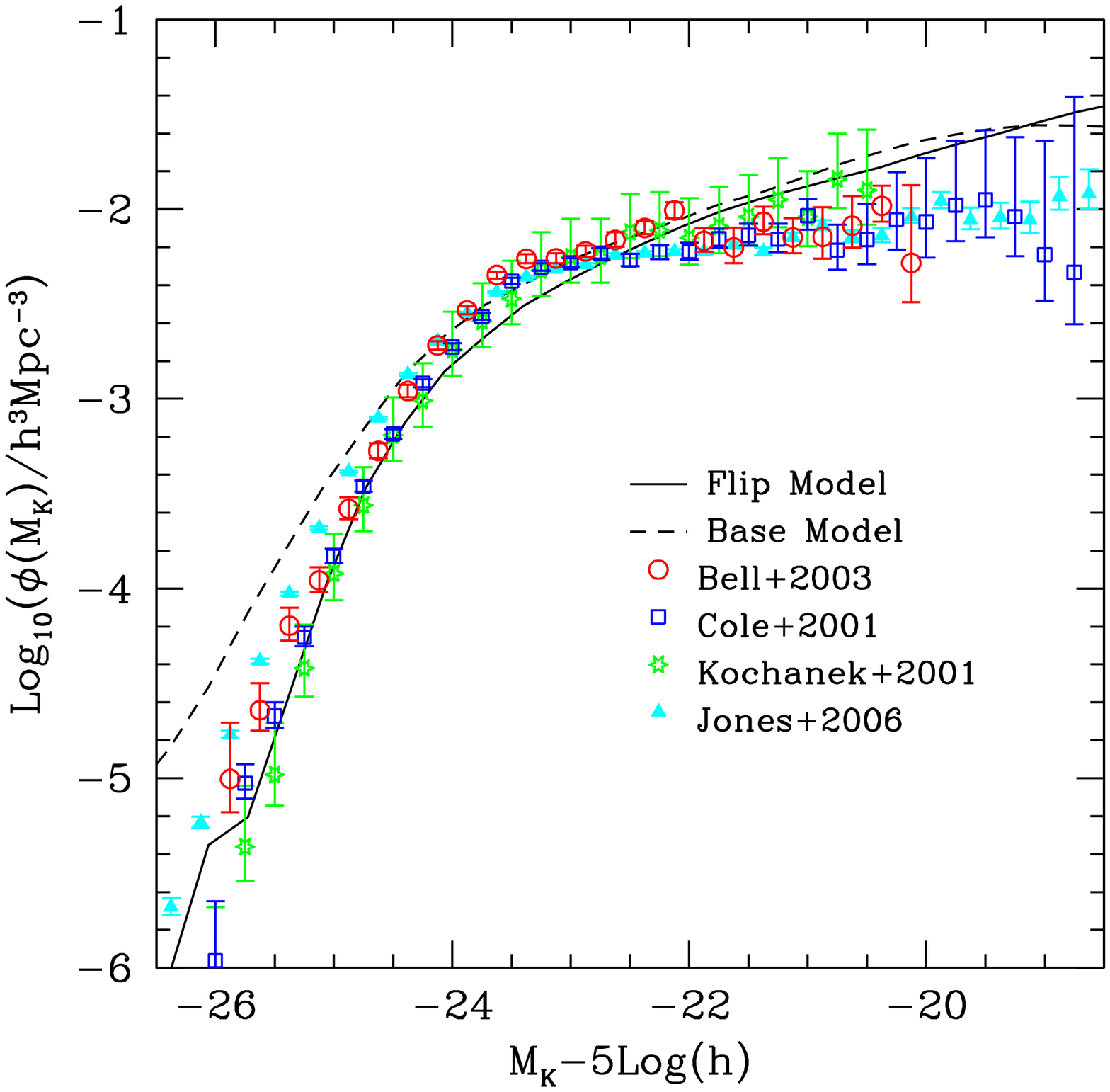}
    \caption{Galaxy luminosity functions in the $b_J$ and $K$ bands (top and bottom panels, respectively), 
which are the most important of
the statistics used in the calibration of 
the semi-analytic model, for the base and flip versions (dashed and solid lines, respectively).  The constraining
observational data are shown as different symbols as stated in the key,
for the estimates by Norberg et al. (2001) and
Jones et al. (2006), both in the $b_J$ band,
and from Bell et al. (2003), Cole et al. (2001), Kochanek et al. (2001) and Jones et al. (2006), 
for the $K$-band.  The luminosity functions correspond to
two versions of the model run with the same parameters obtained for the base model
using the PSO technique (Ruiz et al., 2013).  The semi-analytic
parameters are given in Subsection \ref{sec:comparison}.  \label{fig:lf}}
  \end{center}
\end{figure}

In the first case, when the increase in mass is due to accretion of cool gas, the radius of the disc, $r_{\rm disc}$,  is impeded
from growing and, relative to the no-flip model, becomes slightly smaller
as a result of
the drop in its dimensionless spin parameter by
\begin{equation}
r_{\rm disc}=\eta_{\rm disc}\,r_{\rm disc,no flip}.
\end{equation}
This immediately translates into a smaller dynamical timescale for
the disc with respect to the no-flip model, $t_{\rm dyn,disc}=r_{\rm disc}/V_{\rm vir}$, where $V_{\rm vir}$ 
is the subhalo virial velocity, and a
higher surface density of gas in the disc,
which has a strong impact on the star formation rate of the galaxy,
since the model follows the star formation law given by Croton et al. (2006),
\begin{equation}
\dot m_*=\alpha_{\rm SF}(m_{\rm cold}-m_{\rm crit})/t_{\rm dyn,disc},
\end{equation}
where $\alpha_{\rm SF}$ is the efficiency of star formation, 
a free parameter of the model,
and
$m_{\rm crit}$ is the critical mass for star formation 
given by 
\begin{equation}
m_{\rm crit}=3.8\times 10^9 \left( \frac{V_{\rm vir}}{200 \rm{km s}^{-1}} \right) \left( \frac{r_{\rm disc}}{10\rm{kpc}} \right) M_{\odot}.
\end{equation}
As a result, a flip that impedes the growth of the disc size
increases the surface density of gas in the disc, which in turn
lowers the critical mass for star formation.  This implies 
that galaxies with flips form more stars (at least for equal star formation efficiency parameters).
We adopt a Salpeter initial mass function for all star formation events in the model.

Our model allows for bursts during disc instabilities and mergers.
A galaxy becomes unstable when the quantity
\begin{equation}
\epsilon_{\rm d} \equiv \frac{V_{\rm max}}{(GM_{\rm disc}/r_{\rm disc})^{1/2}}
\end{equation}
is smaller than a critical value of $1$. 
Here, $G$ is the universal gravitational constant and $V_{\rm max}$ is the maximum circular velocity of the disc.
We require the effect of a perturbing galaxy as
a necessary condition to trigger the disc instability.
When a neighbour galaxy
perturbs the unstable disc, all the stars and cold gas in the disc
are transferred to the bulge, and this cold gas is consumed in
a starburst. 
A perturbation triggers a starburst only when the 
typical distance to 
the galaxies sharing the dark matter halo of the unstable one
is smaller than 
$d_{\rm pert}$ times the scale-length of the 
unstable galaxy. 
Smaller galaxy discs are characterised by lower values of $\epsilon_{\rm d}$. They 
will therefore tend to produce more frequent instability events, i.e., 
more frequent bursts of star formation as is seen to occur in hydrodynamical simulations (Sales et al., 2012).
They will also tend to produce shorter lived, episodic discs.
{ Additionally, due to the resolution of the Millennium II simulation used to construct the look up 
tables of slews and flips, the value of $\eta_{\rm disc}$ shows an attractor at small values for a small
fraction of the population.  Therefore, whenever the condition $\eta_{\rm disc}<0.1$ is satisfied, we also
trigger a burst of star formation as in the case of disc instabilities.}

The second case is when the accretion is due to a merger. In the new implementation, we do not apply the
rules for merger driven starbursts following ratios of mass between intervening galaxies as
in other semi-analytic models, or the previous versions of this one (e.g. Cole et al., 2000; Cora 2006; Lagos, Cora \& Padilla, 2008; Tecce et al. 2010). 
We just allow the large
flips produced by mergers to produce highly dense discs, which naturally become unstable and undergo bursts. This is an important change
introduced in the current version of this semi-analytic model.

\subsection{Base model and flip model: comparison of general galaxy properties}
\label{sec:comparison}

The implementation of slews and flips that change the sizes of galaxy discs and
the different criterion adopted to allow starbursts during mergers
will make for a different galaxy population with respect to the base model.
{ In order to ensure that these differences come only from the new implementation,
we will use the same set of model parameters for both models.

We tune the free parameters using the base model.  We allow to vary the}
star formation efficiency $\alpha_{\rm SF}$,
the efficiency of supernovae feedback $\epsilon$, 
the efficiencies of black hole 
growth during bursts (mergers and disc instabilities) and gas cooling,
which are given by  
$f_{\rm BH}$ and $k_{\rm AGN}$, respectively  
(see Lagos, Cora \& Padilla, 2008, for further details),
and the parameter $d_{\rm pert}$ involved in the condition for a perturber
galaxy to trigger a starburst in an unstable galaxy.  
This calibration is done via
the Particle Swarm Optimization (PSO) search 
method (presented in Ruiz et al., 2013),
using the local galaxy luminosity function in the $b_J$ and $K$ bands, and
the relationship between black hole mass and bulge mass 
as observational constraints.
Table \ref{table1} presents
the free parameters and their best fit values found with the
PSO technique for
the base model.
\begin{table}
\centering
\begin{tabular}{cc}
\hline
\hline                                                                                                                                  
Parameter & value  \\
\hline                                                                                                                                  
\hline                                                                                                                                  
$\alpha_{\rm SF}$ & $0.20$ \\
\hline                                                                                                                                  
$\epsilon$ & $0.25$  \\
\hline                              
$f_{\rm BH}$ & $0.026$   \\
\hline                              
$\kappa_{\rm AGN}$ & $1.4\times10^{-3}$  \\
\hline                              
%$f_{\rm bin}$ & $0.046$ & $$ \\
%\hline                                                                                                                                  
$d_{\rm pert}$ & $21.8$ \\
\hline
\hline                                                                                                                                  
\end{tabular}
\caption{Best fit parameters found with the PSO technique for the base model. 
}
\label{table1}
\end{table} 

\begin{figure}
  \begin{center}
    \includegraphics[width=.44\textwidth]{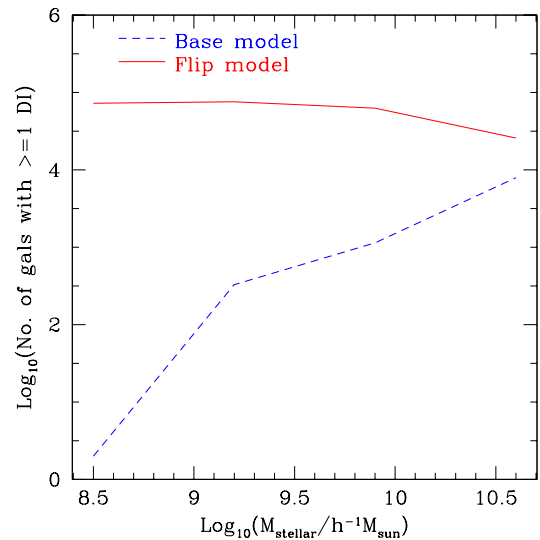}
    \includegraphics[width=.44\textwidth]{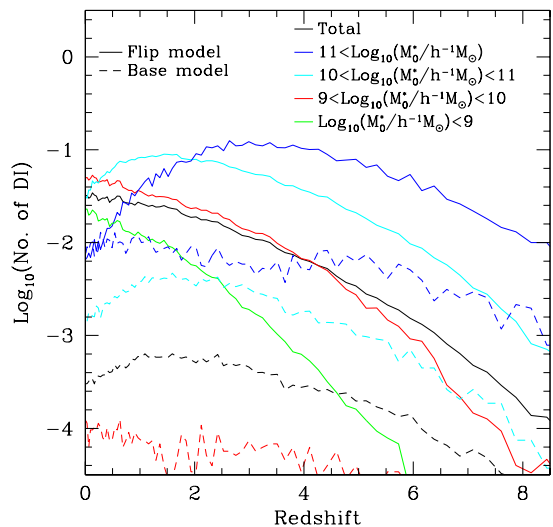}
    \caption{Statistics of the number of DI events in the base and flip models (solid and dashed
lines, respectively). Top panel:
number of galaxies with at least one DI event in their lifetimes, as a function of stellar mass.  Poisson
errors are too small to be noticed for the flip model.
Bottom panel: Number of instabilities per unit redshift suffered by galaxies in different present-day stellar
mass ranges and for the total $z=0$ galaxy population (different colours, as indicated in the key).
\label{fig:ngaldi}}
  \end{center}
\end{figure}

{ Differences between the models are expected as in the flip case 
a higher star formation rate should be produced due to the lower average specific
angular momentum of discs resulting from the flips.
Additionally,
there will probably be a larger number of triggered disc instabilities (cf. Figure \ref{fig:ngaldi}) 
and this, in turn, could provide more material 
for black hole growth during bursts. 
Another expected effect is the increase in supernovae feedback
which could influence the abundance of dwarf galaxies.   These processes, however, 
are deeply intertwined.}

The results for the  
the local galaxy luminosity function in the $b_J$ and $K$ bands, 
which are used as constrains for the search of the best fit parameters
are shown in Figure \ref{fig:lf} for both the base and flip models.  
The latter is much better able to reproduce the knee of the luminosity 
functions in comparison 
to the former.  The flip model also provides a better match to the 
bright end of the luminosity function in both bands,
particularly for the $K$-band, { even when no additional tuning of parameters was performed}.

{ We now turn to compare
the frequency of bursts in the two models}, first concentrating on the disc instability events that affect
the star formation activity and lifetime of discs, and then focusing on mergers which induce
large flips and therefore allow us to use disc instabilities instead of mass fractions to 
trigger starbursts.

\section{Disc instabilities}
\label{sec:di}

The number of discs instabilities (DI) in the two calibrated models are found to be different given the
tendency towards smaller sizes of discs in the flip model.  The upper panel of Figure \ref{fig:ngaldi} shows the number
of galaxies that suffered at least one disc instability event in their lifetimes (excluding instabilities
triggered by mergers in the flip model), as a function of stellar mass.
The solid line corresponds to the flip model in which, as can be seen, more galaxies suffered bursts of star
formation triggered by disc instabilities by even more than two orders of magnitude, more so for
lower mass galaxies.  The lower panel shows this excess
in more detail.  Galaxies with high present-day stellar masses have had more instabilities per unit redshift
in the flip model at all redshifts, and galaxies with different present-day masses all tend 
to reach similar maximum average numbers of DIs per unit redshift.  
In the base model, however, galaxies with low present-day masses suffered very little DIs in comparison
with higher present-day mass galaxies.
The increase in instabilities in the flips versus the base model becomes more important at lower redshifts, with 
differences of 2 orders of magnitude or more.

\begin{figure}
  \begin{center}
    \includegraphics[width=.44\textwidth]{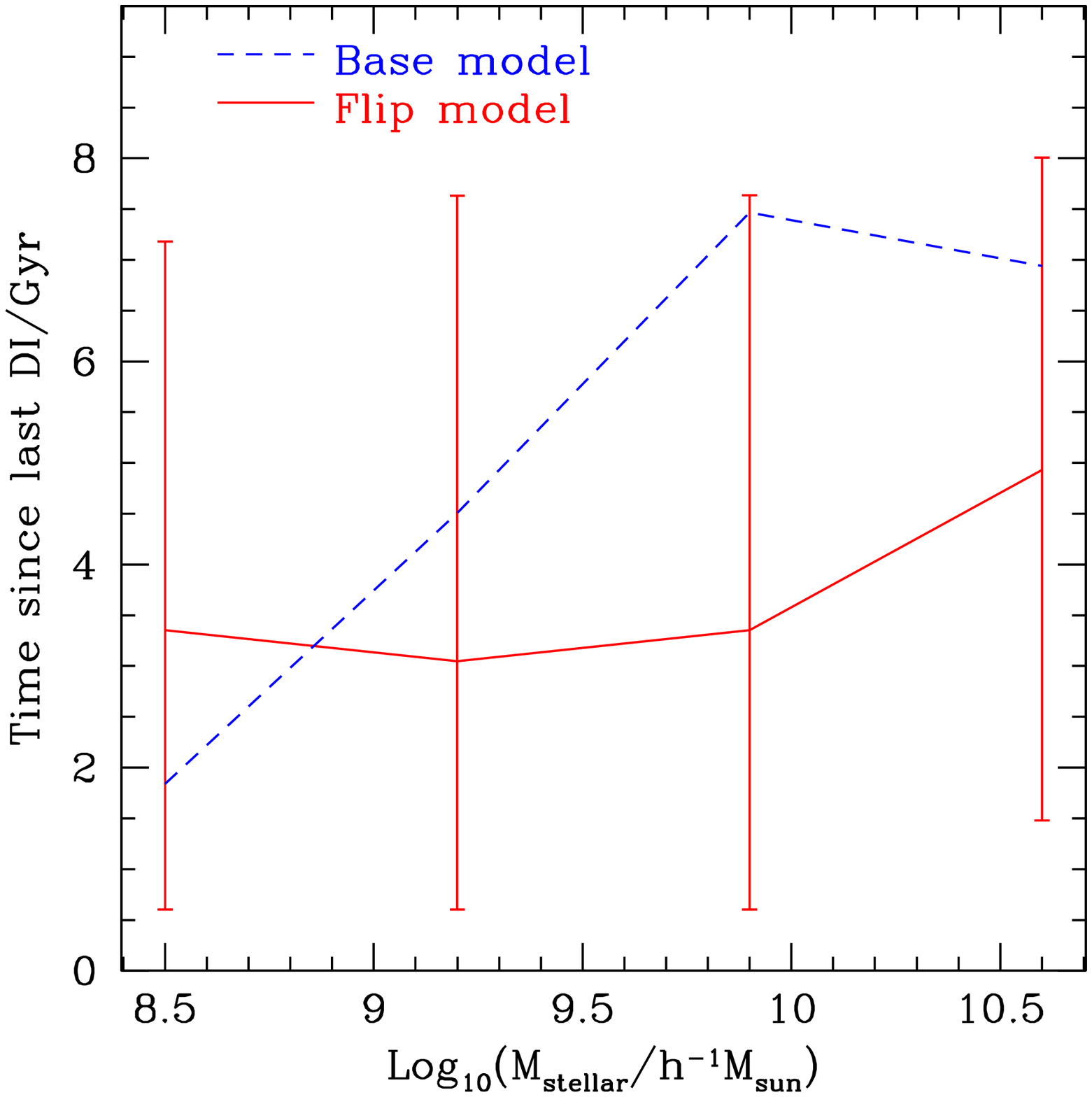}
    \includegraphics[width=.44\textwidth]{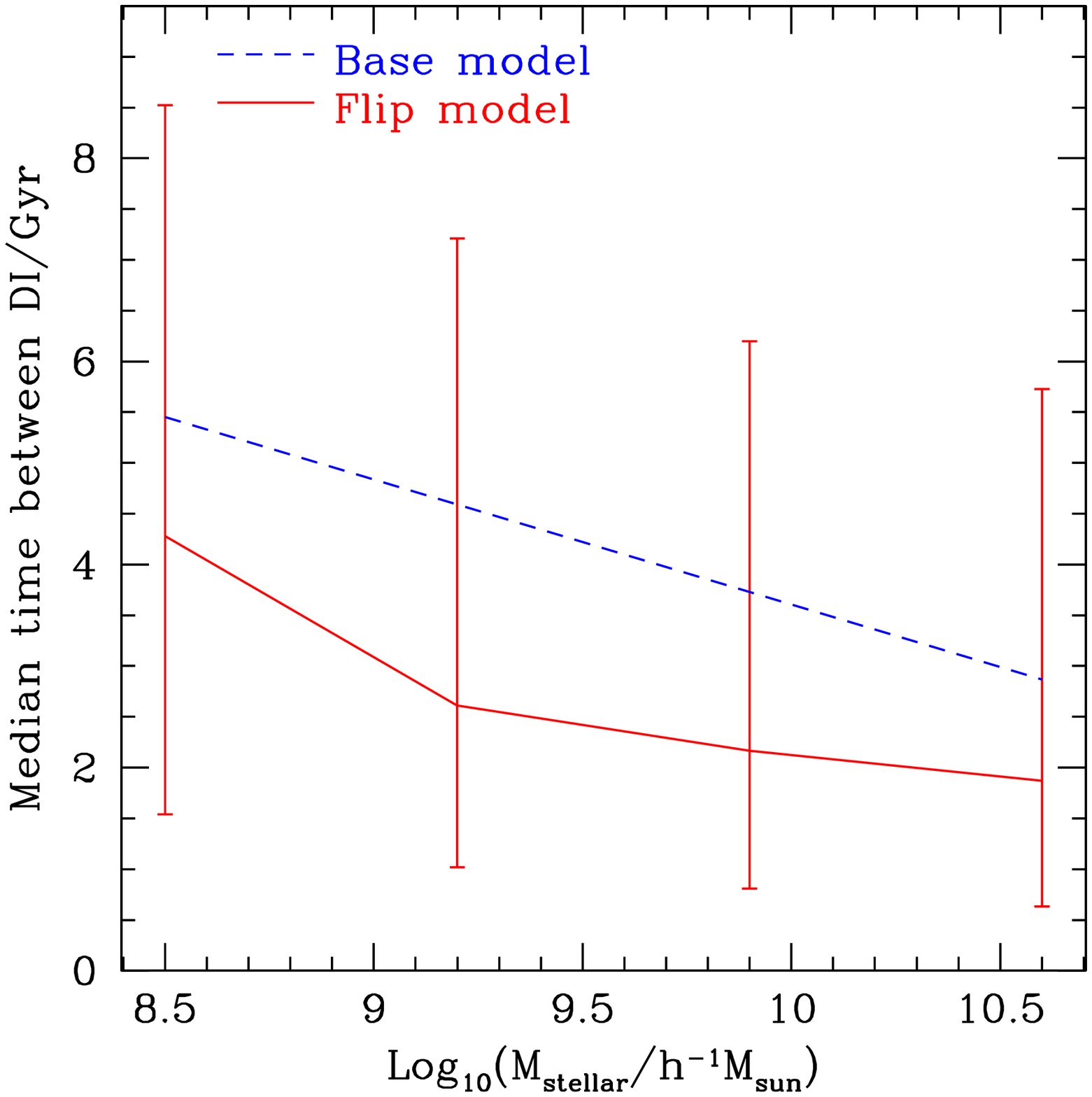}
    \caption{Statistics of disc lifetimes in the base and flip models (dashed and solid lines, 
respectively). Top panel: time since the last disc instability event, which can be interpreted as the
age of present-day discs, as a function of galaxy stellar mass. 
Bottom: median time between consecutive 
disc instabilities, which effectively represents the typical lifetime of discs previous to the
surviving disc (whose age is represented in the top panel), for galaxies that have suffered at least one
disc instability. Errorbars correspond to the $20$ and $80$ percentiles of the distribution,
shown only for the flip model for clarity.\label{fig:di}}
  \end{center}
\end{figure}

The increase in the frequency of disc instabilities which effectively destroy discs
translates into different times of disc survivability.  We explore these changes in Figure \ref{fig:di}, where
the top panel shows the time since the last destruction of a disc by triggered disc instabilities (i.e.
not including merger driven starbursts), as a function of stellar mass.  In this
case we are showing the median for the full galaxy population, so in some cases there may not actually be discs
in the sample, particularly for high stellar mass 
central
galaxies where the AGN feedback is able to prevent a new
disc from forming,  
or for satellite galaxies which do not receive further supply of cool gas.
However, in the cases where a disc is present, this time corresponds to the age of the 
disc that survives to $z=0$.  
As can be seen, surviving discs in the flip model show a trend of 
larger
ages for higher stellar masses, ranging from $\sim 3\,$Gyr to $\sim 5\,$Gyr 
for 
galaxies of $M^*=3\times10^8\,$h$^{-1}\,M_{\odot}$ to
$M^*=10^{11}\,$h$^{-1}\,M_{\odot}$, { with a large scatter that extends up to $\sim 8\,$Gyr. }
The base model shows a similar trend, but surviving discs are roughly a couple of gigayears older than in the
flip model, with a slightly larger difference for higher mass galaxies.
Notice that in both models discs in small galaxies are relatively young, but discs in massive
galaxies are older, as is the case of the Milky Way (e.g. Jimenez et al., 1998, Purcell et al., 2008,
Kazantzidis et al., 2008, Hopkins et al., 2008).  In particular, both models allow the existence
of very old discs for grand-design spiral galaxies.

\begin{figure}
  \begin{center}
    \includegraphics[width=.44\textwidth]{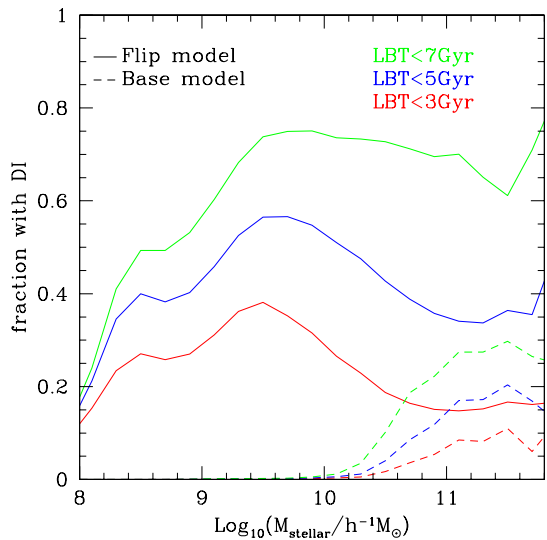}
    \caption{
Fraction of galaxies that have suffered at least one disc instability in the last $3$, $5$, and $7$ gigayears
(red, blue and green lines, respectively), as a function of stellar mass,
for the base and flip models (dashed and solid lines, respectively).
    \label{fig:fracinstab}}
  \end{center}
\end{figure}

The lower panel of Figure \ref{fig:di} shows the average time between consecutive destructions of the disc down to the last
burst driven by disc instabilities, i.e., it does not include the last surviving disc of the upper panel.  Also,
as in the top panel, this statistics does not include merger induced bursts.  
The base model presents a clear trend, where low mass galaxies have a higher average time
between disc destruction events of $5.5\,$Gyr to $3.8\,$Gyr for $M^*=3\times10^8\,$h$^{-1}\,M_{\odot}$ and
$M^*=10^{11}\,$h$^{-1}\,M_{\odot}$, respectively. 
In the flip model the trend is similar, with a change from
$~4.3\,$Gyr to $1.9\,$Gyr for the same stellar masses, but more importantly, the typical disc lifetime is shorter; discs are
more episodic.  This is due, in part, to the fact that slews and flips affect more intensely
galaxies while they are in a fast growth regime;
in this case, the angular momentum flips are more important since the infalling
mass is often a large fraction of the previous disc mass.  Even galaxies that reached high masses at $z=0$ went
through this stage at some point, and therefore show a lower average disc lifetime than in the base model.
 In the latter, on the other hand, disc instabilities take more time to occur
since only when they reach a critical mass they become unstable.
In both models,
high mass galaxies grew in mass during an epoch of faster mass growth (e.g. Lagos, Padilla \& Cora, 2009b)
than lower present-day mass galaxies
and, thus, show lower average times between consecutive destructions of the disc.

A possible shortcoming of the flip model is that the fraction of galaxies that have suffered instabilities
could have grown to an unacceptable level.  
{\bf 
However, we find that the increase in frequency of instabilities
is present mostly at low stellar masses and high look back times (LBT), as can be seen
in Figure \ref{fig:fracinstab}.  For stellar masses $M_{stellar}>10^{11}$h$^{-1}M_{\odot}$ the flip
model shows only about $50$ percent more instabilities per galaxy than the base model for LBT$<3Gyr$.
For the same mass range, for LBT$<5$ and $7Gyr$, the flip model shows about twice as many instabilities as 
the base model.
Notice that a LBT of $3Gyr$ is about one order of magnitude longer than a typical dynamical time,
and therefore present day massive galaxies should show signs of a recent disc instability
in only $\sim 10-15$ percent of the cases, for both models.  The main difference is that in the flip
model, this percentage remains roughly constant with stellar mass down to $M_{stellar}=10^{8}$h$^{-1}M_{\odot}$.
}

\section{Disc instability driven starbursts in mergers}
\label{sec:mergers}

\begin{figure}
  \begin{center}
    \includegraphics[width=.44\textwidth]{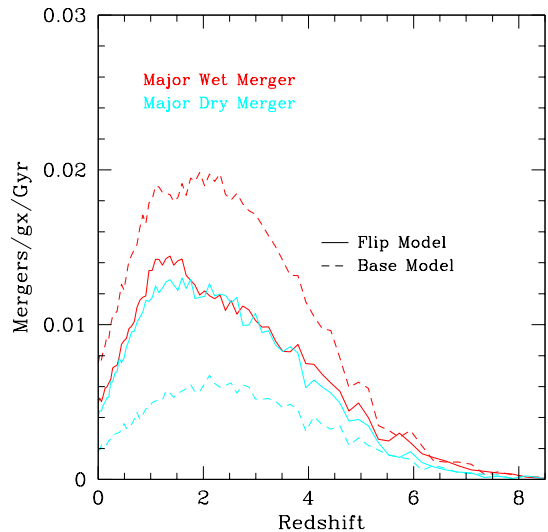}
    \includegraphics[width=.44\textwidth]{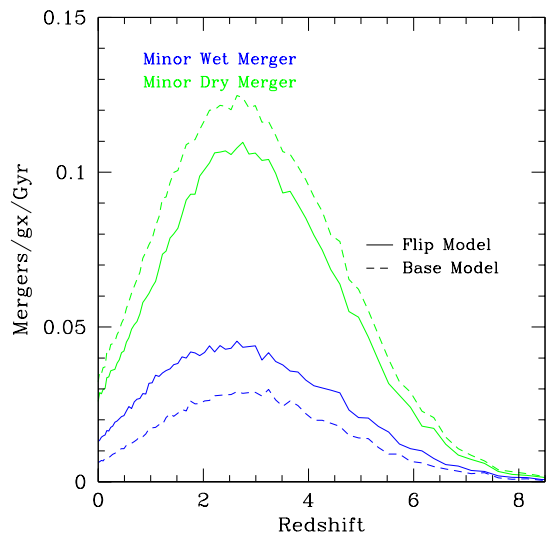}
    \caption{
Frequency of mergers per galaxy per gigayear, separated into events with different mass ratios between
the merging galaxies corresponding to major (mass ratios $>0.3$, top panel) and minor (bottom) mergers.
The flip and base models are represented by solid and dashed lines, respectively. 
Different colours denote wet and dry mergers (see the figure key), where the former correspond to
events where a starburst was triggered, whereas the latter refer to mergers with no ensuing star formation.
    \label{fig:mergers}}
  \end{center}
\end{figure}

As was mentioned in Section \ref{sec:flips}, in the flip model we do not make use of mass ratios
to decide whether a galaxy that has just undergone a merger suffers a burst of star formation; rather we
simply check whether the disc 
of the remnant galaxy
is stable or not.

As the actual mergers of galaxies are governed by the merger trees extracted from the numerical simulation,
both versions, the base and flip models, will have the same total number of mergers between galaxies.
However, since the evolution of the baryons in each model is different, the fraction of mergers
with baryonic mass ratios above a given threshold can be different. 
Furthermore, 
the conditions for the
triggering of bursts are completely different between the models, and therefore the number
of mergers with starbursts are also not expected to be the same.

In the base model, the parameters that are used to decide whether there is a burst in a merger 
are (i) $f_{\rm major}=0.3$, the mass ratio that defines a major merger such
that if two colliding galaxies are characterised by a higher mass ratio, all the cold gas will be
turned into stars and sent to the bulge, along with any other stars in the discs of the merging galaxies;
since in rare cases there will be no cold gas in the colliding galaxies, the fraction $f_{\rm gas,major}=0.6$
is used to define the major merger as dry or wet, depending on whether the fraction of gas in the disc
of the accreted satellite is below or above this threshold (Jim\'enez et al., 2011); notice, however,
that $f_{\rm gas,major}$ is not a parameter of the model as it does not affect the evolution of the galaxies.  (ii)
$f_{\rm disc, burst}=0.6$ is the fraction of gas in the larger of the galaxies involved in a minor merger (i.e. with
a mass ratio lower than $f_{\rm major}$), above which there will be a burst that will transform all the available cold
gas into stars and add them to the bulge,
with the stars in the disc left untouched in this case,  
and (iii)
$f_{\rm burst}=0.05$, the minimum mass ratio for the $f_{\rm disc, burst}$ condition to take place; if this is
not satisfied, there is no burst. The flip model drops three of the parameters of the base model by
simply producing a burst if the disc 
of the remnant galaxy 
is unstable, 
considering the merger itself sufficient enough a perturbation to trigger the instability.
This way the flip model
greatly diminishes the parameter space freedom typical of semi-analytic models.

We compare the frequency of mergers either with or without a starburst in the base and flip models in Figure
\ref{fig:mergers}. We separate the statistics in major and minor mergers according to the ratio of the masses
of the merging galaxies and show the results in the upper and lower panels of the figure, respectively,
for the flip and base models
(solid and dashed lines,
respectively).  The colours help distinguish between
dry and wet mergers.
In the latter type of merger a starburst
takes place, whereas in dry mergers there is no immediate star formation as a result of the merger.
The exception to this rule are dry major mergers in the base model which correspond to 
the case when the satelite cold gas mass fraction is below 
$f_{\rm gas,major}$.

As can be seen, given that the merger trees extracted from the numerical simulation set the total frequency
of mergers, the overall merger rates in both models appear similar at first sight.  
However, there are important differences 
between them.
The flip model presents a larger quantity
of dry major mergers, by about a factor of $2$ compared to the base model.  
Given that a major merger produces an important flip in the angular momentum of a galaxy
disc, and therefore impedes the growth of the size of the galaxy, the fact that there are no bursts in these cases 
is an indication that 
the remnant galaxy is not massive enough to become unstable and thus trigger a starburst.
In the base model, any available gas will undergo a burst. 
However, as was shown in the previous section, in the flip model
the frequency of disc instabilities is much higher than in the base model, which
acts to balance the offset in merger driven starbursts in the model, causing little
difference to the galaxy colours and to the morphological fractions as a function of stellar
mass, for instance {(cf. following Section and Figure \ref{fig:morph}).}

{ In the case of minor mergers, the results are very similar between the two models,
with only a $15$ percent deficit (increment) of dry (wet) minor mergers in the flip with respect to the
base model.
Another difference with respect to the major merger case is then that
this slight} excess of minor wet mergers is not counteracted by the bursts driven by DIs (cf. previous section), since both effects
go in the same direction of having more bursts when the accretion of baryonic matter carries
small amounts of matter.  

In the following sections we will then use the flip model with disc instabilities as the only driver of 
starbursts in galaxies,
both in perturbed galaxies and in the remnant galaxies of mergers,
 bearing in mind the possible influence of more frequent bursts in smooth accretion
events and minor mergers.

\section{Testable consequences}
\label{sec:results}

Even though the $z=0$ galaxy populations in the base and flip models show similar luminosity functions (Figure \ref{fig:lf}),
it is clear that there are differences in other galaxy properties. In this section we show some of these differences
and make further comparisons to observational data, when available.

\begin{figure}
  \begin{center}
    \includegraphics[width=.44\textwidth]{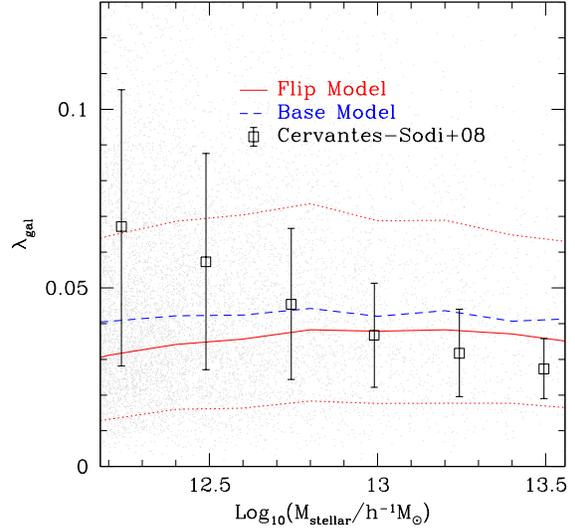}
    \caption{
Dimensionless spin parameter of galaxies as a function of their host dark matter halo mass.  The open squares with errorbars correspond to measurements for SDSS galaxies by Cervantes-Sodi et al. (2008).  The grey points show the scatter of galaxy spins in the flip model; the solid red line corresponds to the median of these points and the dotted lines show the $10$ and $90$ percentiles.  The blue dashed line shows the median of the base model.  
    \label{fig:spin}}
  \end{center}
\end{figure}

{ Our first comparison is focused on the dimensionless spin parameter of galaxies.  Cervantes-Sodi 
et al. (2008) measured this quantity for a sample of 
galaxies taken from the Sloan Digital Sky Survey (SDSS, York et al., 2000) using a simple model for
their dynamical structure.  
Figure \ref{fig:spin} shows their results compared to our models.  As can be seen, 
when assuming the spin of the galaxy to coincide with that of the disc, then both models
provide similar agreement, but without the trend of higher spins for lower stellar masses.  The flip
model shows slightly lower spins due to 
the effect of the slews and flips.
 }

\begin{figure}
  \begin{center}
    \includegraphics[width=.44\textwidth]{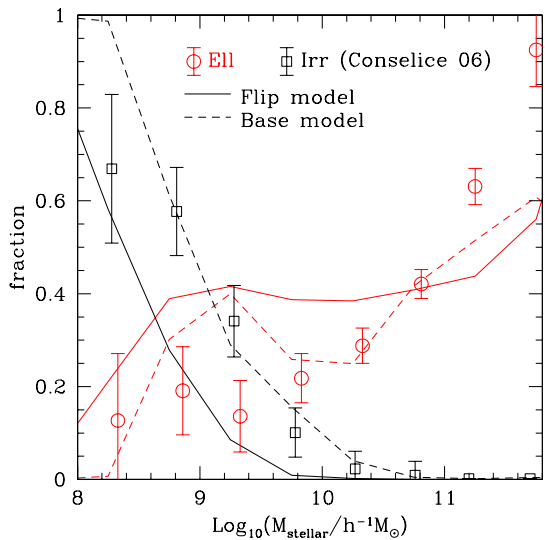}
    \caption{
Fractions of galaxies with elliptical and irregular morphology (red and black, respectively), 
as a function of stellar mass, for the base (dashed lines) and flip (solid lines) models.  The symbols with
errorbars show the results for nearby galaxies obtained by Conselice (2006) for the same galaxy types.
    \label{fig:morph}}
  \end{center}
\end{figure}

The change in the frequency of disc instabilities, and the mechanism to trigger starbursts in
general, can cause the morphological fractions to vary between
the base and flip models.  Therefore, we compare the $z=0$ fractions of galaxies with different
morphologies as a function of their stellar mass in Figure \ref{fig:morph}.  Model galaxies
are separated into different morphological classes according to their bulge to total stellar
mass ratios. {\bf Irregular galaxies are those with no bulge,}
\footnote{\bf The results of Figure \ref{fig:morph} do not change significantly when changing
the upper limit in bulge-to-total stellar mass ratio for irregulars from $0$ to $5$ percent.
} 
spirals contain a bulge which can make up
for up to $80$ percent of the total stellar mass, and ellipticals contain more than $80$  percent of their
stellar mass in a bulge.  Since the total population of galaxies is divided into one of these three
types, we only show the results for irregular and elliptical galaxies.  {\bf As can be seen,
the base model shows good agreement
for the elliptical and irregular galaxies, with only a slight
excess of ellipticals around $M_{stellar}=10^{9}$h$^{-1}M_{\odot}$ and of irregulars at
$M_{stellar}=10^{8}$h$^{-1}M_{\odot}$.  In the case of the flip model,
the higher frequency of disc instabilities is responsible for a drop in the frequency of galaxies with
no bulge, which is only slightly at odds with the data and at the lowest stellar masses the agreement 
improves.  
The flip model shows reasonable agreement with the observed fraction of elliptical galaxies, although
slightly less so than the base model.  Notice though that the flip model parameters were
not recalibrated to fit any observables, and the agreement which is already reasonable 
could improve further with a slightly different
set of parameters.  Also, we remind the reader that the flip model does qualitatively improve the agreement with 
the knee and bright end of the luminosity functions (cf. Figure \ref{fig:lf}); these tests simply show that it
also provides a reasonable frequency of bulges.
}

\begin{figure}
  \begin{center}
    \includegraphics[width=.44\textwidth]{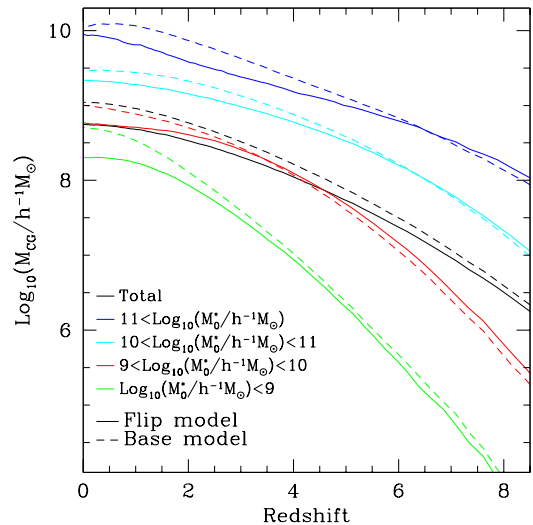}
    \includegraphics[width=.44\textwidth]{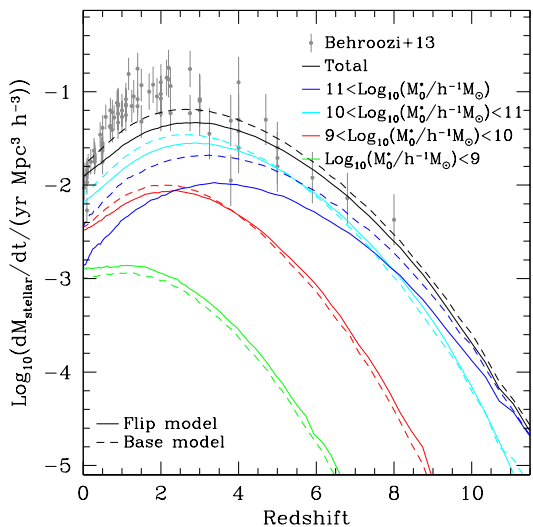}
    \caption{
Top panel: cold gas mass for all the $z=0$ galaxies in each model (black) and different present-day stellar mass ranges (colours,
see the figure key), for the base and flip models (dashed and solid lines, respectively).  Bottom: star formation rate
density as a function of redshift for the full $z=0$ galaxy population (black) and different present-day stellar mass
ranges for the base and flip models (same colours and line types as top panel).  The grey points with errorbars correspond
to the observational compilation by Behroozi et al. (2013). 
    \label{fig:cgsfr}}
  \end{center}
\end{figure}

Figure \ref{fig:cgsfr} shows a comparison 
of the cold gas mass and star formation rate density as a function of redshift (top
and bottom panels, respectively) resulting from the two models, for the full galaxy population and for galaxies with different present-day stellar masses, as
indicated in the figure key.  The star formation rate densities of the models are further compared to
the data compiled by Behroozi et al. (2013).  The model star formation rate densities were divided by $1.515$ to convert
them to the Chabrier initial mass function 
adopted by Behroozi et al..
In both panels, the solid lines represent the flip model, whereas the dashed lines correspond
to the base model.  As expected, due to the lower star formation activity in the latter, the amount of cold gas in 
galaxies is higher in the base model, particularly at low redshifts; 
the flip model spends more cold gas as a result of the slower growth of angular momentum due to the
action of
the slews and flips.  In terms of the full galaxy population, 
the different models differ in this quantity by about a factor of $2$ at $z=0$.

{ The lower panel of the figure shows that at high redshifts the star formation rate
density tends to be higher in the flip model, except for the most massive galaxies.  
Then, at lower redshifts for lower final stellar mass galaxies, the star formation rate
densities of the two models equal each other until the base model starts to show higher rates.
For instance, this transition occurs at $z \sim 4$ for  $10^{9}\,$h$^{-1}\,M_{\odot}<M^*<10^{10}\,$h$^{-1}\,M_{\odot}$,
and at $z \sim 5.5$ for  $10^{10}\,$h$^{-1}\,M_{\odot}<M^*<10^{11}\,$h$^{-1}\,M_{\odot}$, whereas
for 
$M^*<10^{9}\,$h$^{-1}\,M_{\odot}$ the star formation activity is still higher for the flip model at $z=0$.
This effectively shows that the flip model is downsizing more effectively than the base model.}
The grey symbols represent the compilation of observational results by Behroozi et al. (2013).  As can be seen,
the scatter in the observations is similar to that present between the base and flip models, making both of them
roughly consistent with the observations.

\begin{figure}
  \begin{center}
    \includegraphics[width=.44\textwidth]{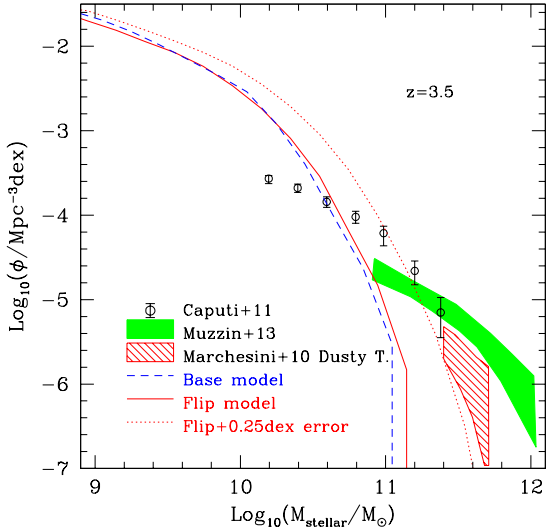}
    \caption{
Stellar mass functions for model galaxies at $z=3.5$ for the base and flip models (dashed and solid lines,
respectively).  The dotted line shows the results for the flip model if a gaussian error of width $0.25$dex
in the logarithm of the stellar mass is added to the model masses.  
The shaded regions correspond to the NEWFIRM medium band survey sample of massive galaxies
at $3<z<4$, when adopting the dusty template set (Marchesini et al., 2010), and the results for
the ULTRAVISTA survey by Muzzin et al. (2013), whereas the symbols show the results by Caputi et al.
(2011) for the UDS, as indicated in the key.
\label{fig:smf}}
  \end{center}
\end{figure}

An immediate consequence of having a higher star formation rate at high redshifts
is that the stellar mass in galaxies grows at a faster rate.  Therefore an interesting comparison with observations is
the one that focuses on the stellar mass function which, at high redshifts,
has been difficult to
reproduce by semi-analytic models (e.g. de Lucia et al., 2006),
as was mentioned in the Introduction.
Figure \ref{fig:smf} shows the stellar mass
function for galaxies at $z=3.5$ extracted from the base and flip models (dashed and solid lines, respectively).
As expected, the flip model produces more high stellar mass galaxies, with a general shift of the stellar mass function
toward higher masses by 
$20$ percent.  This increase shortens the gap between the
flip model stellar masses and the estimates from the NEWFIRM medium band survey by Marchesini et al. (2010)
for massive galaxies at $3<z<4$, particularly for the case when they consider dusty templates to fit their
observed photometry.  
In their paper, Marchesini et al. show that this mass function would only be consistent
with the Somerville et al. (2008) galaxies if they included a $0.25$ dex error in the mass estimates for
the model galaxies.  In the flip model case, a $0.25$ dex error (dotted line) makes the stellar mass function agree
with the dusty-template case in Marchesini et al.

The figure also shows a lack of agreement between the flip model and the Muzzin et al. (2013) data 
at the high mass end of the stellar mass function, but it should be taken into consideration that
they did not include a dusty template as was done by Marchesini et al.  In their appendix, Muzzin et al. state that
the number density of the most massive galaxies diminishes considerably when they add this template to their set which
would improve our agreement with their results.
Notice also that the number density of dwarf galaxies with $M_*<10^{10}$h$^{-1}M_{\odot}$ does not change
between the two flavours of this model implying that even though the formalism adopted in the
flip model increases the stellar mass growth at high redshifts, it does not do so at the expense
of a larger population of dwarf galaxies (cf. Henriques et al. 2013).
The dwarfs with stellar masses around the $10^{10}$h$^{-1}$M$_{\odot}$ 
mark in our model appear to be over-abundant in comparison to the results by Caputi et al. (2011)
for the UDS (Williams et al., 2009).
Notice,
however, that Caputi et al. also do not include dusty templates as Marchesini et al.

\begin{figure}
  \begin{center}
    \includegraphics[width=.42\textwidth]{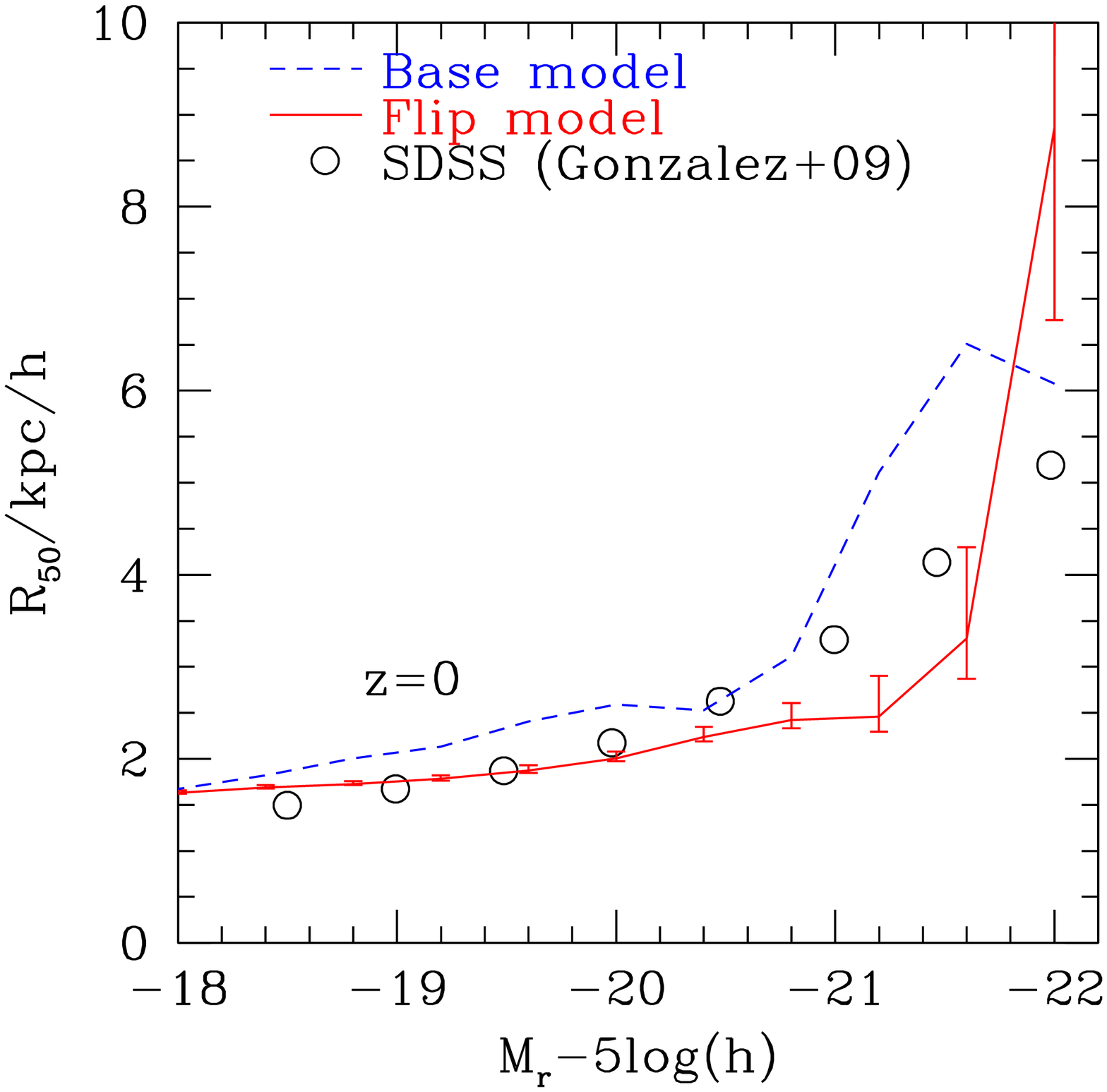}
    \includegraphics[width=.42\textwidth]{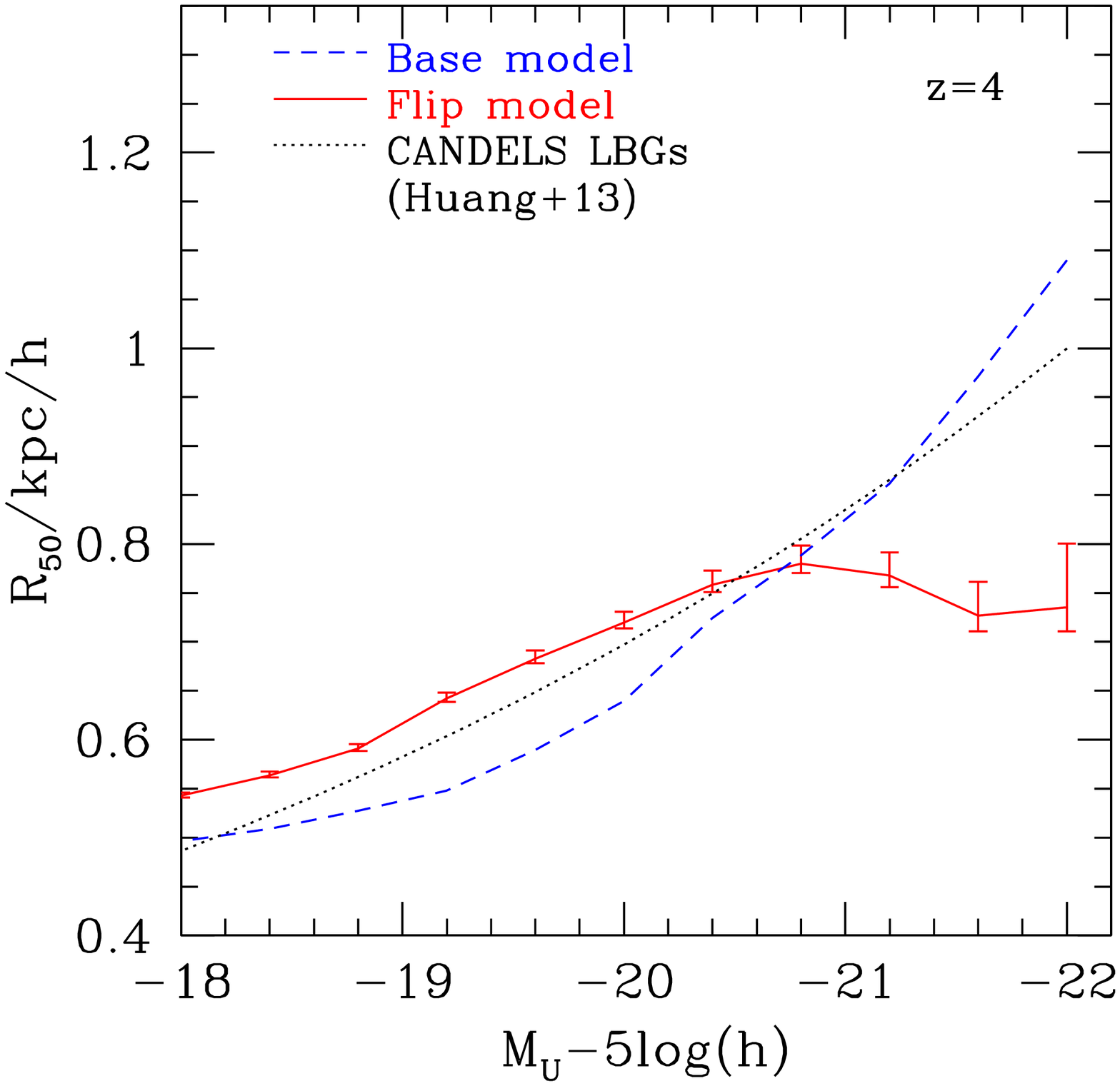}
    \caption{
Galaxy radii containing $50$ percent of the total flux, as a function of absolute magnitude, for the base and
flip models (dashed and solid lines, respectively).  Top panel: $z=0$ results for spirals, as a function of r-band
absolute magnitude, compared to results from the Sloan Digital Sky Survey Main Galaxy Sample from
Gonz\'alez et al. (2009).  Bottom: $z=4$ sizes of star forming galaxies in the models, and Lyman Break Galaxies
 (dotted lines, Huang et al., 2013)
from the CANDELS survey (Wuyts et al., 2011).
The errorbars are shown only for the flip model, and indicate the error on the mean.
\label{fig:radios}}
  \end{center}
\end{figure}

Given the changes to the galaxy sizes imprinted by the inhibition of angular momentum growth due
to flips, it is also necessary to study the distributions of galaxy sizes in the new model.  In principle,
one can think that inhibiting the growth of the angular momentum would
result in smaller disc sizes.  However, when measuring the median size of surviving spiral galaxies at $z=0$
and $z=4$ in the upper and lower panels of Figure \ref{fig:radios}, respectively, we see that this is the
case, but to a small extent.
At $z=0$, the flip model shows similar median galaxy sizes than the base model
almost throughout the entire range of magnitudes shown, and both models are in similar agreement
with SDSS
galaxy sizes measured by Gonz\'alez et al. (2009).  As was shown by Bruce et al. (2012), the spiral disc sizes
at $z=2$ are quite similar to those of $z=0$, and therefore we go to higher redshifts for our next comparison.
We use the recent measurement of $z=4$ Lyman-Break Galaxies (LBGs)
with sizes estimated from CANDELS (Wuyts et al., 2011) data by Huang et al. (2013).  In this case, the base model shows
sizes consistent with the observed ones, and this is also the case for the flip model, except at the bright end where
flip model galaxies flatten their size and become slightly ($25$ percent) smaller than the observed ones.

The reason for the similar sizes between the two models at all redshifts is that 
the flip model first tends to distort the distribution of galaxy sizes by extending a tail
toward small disc radii, but at the same time the disc instability mechanism quickly destroys the smaller discs, making this distribution
lose its small radii tail 
and, consequently, 
show a similar median value to that of the base model.
This effect is present at all luminosities at both redshift ranges.

\section{Conclusions}
\label{sec:conclusions}

In this work we presented the implementation of angular momentum slews and flips in a semi-analytic galaxy formation model.
The aim in doing this was to reproduce in a more realistic way the fact that often the angular momentum of discs
and that of the matter they accrete are not aligned.  This lack of alignment is thought to be responsible for the
destruction of discs within dark matter haloes that do not necessarily undergo a dramatic event such as a merger.
The interaction of
the infalling material with
the disc as it arrives produces the discs to become unstable, undergo bursts, and feed new stars to a bulge.

From the hydrodynamical side, Sales et al. (2012) used the GIMIC galaxies to show that most of the discs that
survive at $z=0$ show angular momentum vectors that are better aligned with the matter at turn around than particles
in a spheroid component.  
This indicates that misaligned infall could be responsible for the formation of bulges.
A possible reason behind this is that the dark matter halo that hosts a galaxy suffers several stochastic 
changes in its angular momentum vector as shown by Bett \& Frenk (2012).  If 
the hot gas that cools toward the disc is relaxed in this dark matter
halo, in most cases the cooling gas will have an angular momentum that will not necessarily be parallel to that
of the cold disc it is destined to feed.

In most semi-analytic models, the resolution of the parent dark matter simulation does not allow for the direction
of the angular momentum of the halo to be measured with high accuracy.  For this, at least $1000$ particles
per halo are needed, and most semi-analytic models use haloes with of the order of $10$ or more particles.
Therefore, we measured the probability of angular momentum flips by a given angle using subhaloes of the
Millennium II simulation (Boylan-Kolchin et al., 2009) after they change their mass by a certain relative amount, in 
different redshift ranges.  We construct different probability distributions for when the accretion is 
smooth or due to mergers.  In particular, we find that at high
redshifts the same relative increase in mass produces larger flips in the angular momentum of haloes.  This is also
the case for mergers in comparison to smooth accretion.

We apply flips to the semi-analytic model {\small SAG} (Cora 2006;
Lagos, Cora \& Padilla, 2008, Lagos, Padilla \& Cora, 2009; Tecce et al. 2010) by using Monte-Carlo 
simulations for the flips
suffered by the galaxy angular momenta.  To do this, we assume that the distributions of flips measured in the subhaloes
apply as well to the baryons at their centres and that, in addition, their effect is that of slowing down the growth of angular momentum 
of discs with
respect to their host haloes. The motivation
behind this assumption comes from the results of hydrodynamical simulations that show that infalling material with
misaligned angular momenta with respect to the galaxy disc could be responsible for its repeated destruction (Sales et al., 2012,
Aumer et al., 2013).
{
In our new implementation, the net effect of the flips is that the average disc slows down its angular momentum acquisition during its most active
growth phase to then continue growing at later times.  The discs that are more affected by this are those
which accrete mass with misaligned angular momentum, and are those more prone to undergo instabilities and
to be added to their bulge.}

Even though discs are more unstable in the flip model, the galaxies that reach high present-day
stellar masses have not been subject to disc instabilities in just as long a time 
as in a model with no flips, i.e.,
grand-design spiral discs are also old in the flip model.  However, the past history of these galaxies includes 
shorter lived previous incarnations of discs, i.e. more episodic discs, 
since the flips are more important while a galaxy is still small
as it is growing in size in larger relative fractions of its mass compared to higher mass galaxies.

{ 
As the inclusion of flips increases the star formation activity at high redshifts, the stellar mass function
of high redshift galaxies reaches slightly higher masses slightly improving the agreement 
with observations at high redshift.

}

We also take advantage of the large flips suffered by galaxies during mergers to use the disc instability condition 
to produce bursts in mergers.  This way our model reduces its parameter space, dispensing of three parameters that,
in the base model, are used to determine whether there are bursts via conditions on different mass ratios of the merging galaxies.  
The resulting fractions of wet mergers (i.e. with bursts of star formation) in the flip model are reduced 
compared to the base model, but the higher frequency of bursts due to other disc instability events in the former
compensates for this and makes a final galaxy population that is similar in the two models in terms
of colours and morphologies.  

{\bf Massive galaxies in the flip model form stars more rapidly at early times, and have less available cold 
gas for star formation
at later times providing a better fit to the steep fall off of the bright end of the luminosity function.}

\section*{Acknowledgements}

This work was supported in part by Fondecyt Regular No. 1110328, BASAL PFB-06 "Centro de Astronomia y Tecnologias Afines".
NP and SC acknowledge support by the European Commissions Framework Programme 7, through the Marie Curie International
Research Staff Exchange Scheme LACEGAL (PIRSES-GA-2010-269264).  We have benefitted from
fruitful discussions with Yuval Birnboim, David Wilman, Bruno Henriques,
Francoise Combes, Laura Sales, Noam Libeskind, Simon White and Carlos Frenk.  We thank the anonymous
referee for helpful comments.  NP thanks the hospitality of the Max Planck
Institute for Astrophysics at Garching, where part of this work was done and several helpful discussions were held.
The Geryon cluster at the Centro de Astro-Ingenieria UC was extensively used for the calculations performed in this paper.
The Anillo ACT-86, FONDEQUIP AIC-57, and QUIMAL 130008 provided funding for several improvements to the Geryon cluster.
This work was partially supported by the Consejo Nacional de Investigaciones Cient\'{\i}ficas y
T\'ecnicas (CONICET, Argentina), Secretar\'{\i}a de Ciencia y Tecnolog\'{\i}a de la Universidad
Nacional de C\'ordoba (SeCyT-UNC, Argentina),
Universidad Nacional de La Plata (UNLP, Argentina), and
Instituto de Astrof\'{\i}sica de
La Plata (IALP, Argentina).
SAC acknowledges grants from CONICET (PIP-220), Argentina,
Agencia Nacional
de Promoci\'on Cient\'{\i}fica y Tecnol\'ogica (PICT-2008-0627), Argentina,
and Fondecyt No. 1110328, Chile.
The Millennium II Simulation database used in this paper and the web application providing online access to
them were constructed as part of the activities of the German Astrophysical Virtual Observatory (GAVO).

\label{lastpage}


\begin{thebibliography}{99}
\bibitem[]{3a12} Algorry, D., Navarro, J., Abadi, M., Sales, L., Steinmetz, M., Piontek, F., 2013, MNRAS, arXiv:1311.1215
\bibitem[]{b312} Aumer, M., White, S., Naab, T., Scannapieco, C., 2013, arXiv:1304.1559 
\bibitem[]{Ua} Baugh, C., 2006, RPPh, 69, 3101
\bibitem[]{Uaa} Behroozi, P., Wechsler, R., Conroy, C., 2013, ApJ, 770, 57
\bibitem[]{Usda} Bell, E., McIntosh, D., Katz, N., Weinberg, M., 2003, ApJS, 149, 289
\bibitem[]{Uf} Benson, A., 2010, PhR, 495,33
\bibitem[]{Ufa} Berry, M, Somerville, R., Haas, M., Gawiser, E., Maller, A., Popping, G., Trager, S., 2013, submitted to MNRAS, arXiv:1308.2598
\bibitem[]{bf12} Bett, P., Frenk, C., 2012, MNRAS, 420, 3324
\bibitem[]{bf13} Bird, J.,  Kazantzidis, S., Weinberg, D., Guedes, J., Callegari, S., Mayer, L., Madau, P., 2013, ApJ, 773, 438
\bibitem[]{cf13} Bois, M., et al., 2011, MNRAS, 416, 1654
\bibitem[]{U523} Bower, R., Benson, A., Crain, R., 2012, MNRAS, 422, 2816
\bibitem[]{U23} Boylan-Kolchin, M., Springel, V., White, S., Jenkins, A., Lemson, G., 2009, MNRAS, 398, 1150
\bibitem[]{31ss} Brook, C. B., et al. 2011, MNRAS, 415, 1051
\bibitem[]{31} Bruce V., et al., 2012, MNRAS,427, 1666
\bibitem[]{31as} Bullock, J., et al., 2001, MNRAS, 321, 559
\bibitem[]{ltravista2} Caputi, K. I., Cirasuolo, M., Dunlop, J. S., McLure, R. J., Farrah, D., \& Almaini, O. 2011, MNRAS, 413, 162
%\bibitem[]{travista2} Cervantes-Sodi, B., Hernandez, X., Park, C., Kim, J., 2008, MNRAS, 388, 863
\bibitem[]{Uhq} Cole, S., Lacey, C., Baugh, C., Frenk, C., 2000, MNRAS, 319, 168
\bibitem[]{Uq} Cole, S., et al., 2001, MNRAS, 326, 255
\bibitem[]{Uqa} Conselice, J.C., 2006, MNRAS, 373, 1389
\bibitem[]{asdhq} Cora, S., 2006, MNRAS, 368, 1540 
\bibitem[]{U134} Crain, R., et al., 2009, MNRAS, 399, 1773
\bibitem[]{U1134} Croton, D., et al., 2006, MNRAS, 367, 864
\bibitem[]{1134} De Lucia, G., Springel, V., White, S., Croton, D., Kauffmann, G., 2006, MNRAS, 366, 499
\bibitem[]{1634} D'Onghia, E., Burkert, A., 2004, ApJ, 612, 13
\bibitem[]{1234} Dubinski, J. \& Chakrabarty, D. 2009, ApJ, 703, 2068
\bibitem[]{1334} Dutton, A., van den Bosch, F., 2012, MNRAS, 421, 608
\bibitem[]{852} Ferreras I., Lisker T., Pasquali A., Khochfar S., Kaviraj S., 2009, MNRAS, 396, 1573
\bibitem[]{8521} García-Ruiz, I., Kuijken, K. \& Dubinski. J. 2002, MNRAS, 337, 459
\bibitem[]{Musddsnichsamguo} Gonzalez, J., Lacey, C., Baugh, C., Frenk, C., Benson, A., 2009, MNRAS, 397, 1254
\bibitem[]{Musnichsamguo} Guedes, J., Callegari, S., Madau, P., Mayer, L., 2011, ApJ, 742, 438
\bibitem[]{Munichsamguo} Guo, X. et al., 2011, MNRAS, 413, 101
\bibitem[]{Munichsam} 	Henriques, B., White, S., Thomas, P., Angulo, R., Guo, Q., Lemson, G., Springel, V., 2013, MNRAS, 431, 3373
\bibitem[]{4} Hopkins P., et al. 2008, ApJ 688, 757
\bibitem[]{4a} Hopkins, P. F., Quataert, E., Murray, N. 2011, MNRAS, 417, 950
\bibitem[]{42s} Huang, K.-H., Ferguson, H., Ravindranath, S., Su, J., 2013, ApJ, 765, 68 
\bibitem[]{123} Jarosik, N., et al., 2011, ApJS, 192, 14
\bibitem[]{1} Jimenez R., et al. 1998, MNRAS 299, 515
\bibitem[]{1432} Jimenez N., Cora, S., Bassino, L., Tecce, T., Smith Castelli, A., 2011, MNRAS, 417, 785
\bibitem[]{11} Jones, D., Peterson, B., Colless, M., Saunders, W., 2006, MNRAS, 369, 25
\bibitem[]{e2} Kannan, R., Stinson, G. S., Macciò, A. V., Brook, C., Weinmann,
S. M., Wadsley, J., \& Couchman, H. M. P. 2013, MNRAS, in
press, arxiv:1302.2618
\bibitem[]{e1} Kauffmann, G, Colberg, J, Diaferio, A., White, S., 1999, MNRAS, 303, 188
\bibitem[]{3} Kazantzidis S., et al. 2008, ApJ 688, 254
\bibitem[]{12} Kennicutt, R., 1998, ApJ, 498, 541
\bibitem[]{12s} Kochanek, C., et al., 2001, ApJ, 560, 566
\bibitem[]{easfd} Lagos, C., Cora, S., Padilla, N., 2008, MNRAS, 388, 587
\bibitem[]{easfd} Lagos, C., Padilla, N., Cora, S., 2009a, MNRAS, 395, 625
\bibitem[]{easfd} Lagos, C., Padilla, N., Cora, S., 2009b, MNRAS, 397, 31L
\bibitem[]{lagos} Lagos C.D.P., Padilla N., Strauss M., Cora S., Hao L., 2011, MNRAS, 414, 2148
\bibitem[]{lag} Maller, A., Dekel, A., 2002, MNRAS, 335, 487
\bibitem[]{1030} Marchesini D. et al., 2010, ApJ, 725, 1277
\bibitem[]{U6} Marinacci F., Pakmor R. \& Springel, V. 2013, arXiv1305.5360
\bibitem[]{ultravista2} McCracken et al. 2012, A\&A, 544, A156.
\bibitem[]{utravista2} Mo, H., Mao, S., White, S., 1998, MNRAS, 295, 319
\bibitem[]{ultravista} Muzzin, A. et al., 2013, arXiv:1303.4409
\bibitem[]{10002} Naab, T., et al. 2013, MNRAS, arXiv:1311.0284
\bibitem[]{1002} Nagamine, K. 2010, Advances in Astronomy, 2010
\bibitem[]{2} Norberg, P., et al. 2001, 328, 64
\bibitem[]{203} P\'erez-Gonz\'alez P., et al., 2008, ApJ, 675, 234
\bibitem[]{20s3} Powell, L., Slyz, A., Devriendt, J., 2011, MNRAS, 414, 3671
\bibitem[]{2} Purcell, C., et al. 2008, arXiv:0810.2785
\bibitem[]{aaa2} Robertson, B., Bullock, J., Cox, T., Di Matteo, T., Hernquist, L., Springel, V., Yoshida, N., 2006, ApJ, 645, 986
\bibitem[]{aaaa2} Romanowsky, A., \& Fall, M., 2012, ApJSS, 203, 17
\bibitem[]{asdfg1} Ruiz, A., Cora, S., Padilla, N., Dominguez, M., Tecce, T., Orsi, A., Yaryura, Y., Lambas, D., Gargiulo, I., Munoz, A., 2013, submitted to MNRAS, arXiv:1310.7034
\bibitem[]{asaf} Saha, K., Naab, T., 2013, MNRAS, 434, 1287
\bibitem[]{asf} Sales, L., Navarro, J., Theuns, T., Schaye, J., White, S., Frenk, C., Crain, R., Dalla Vecchia, C., 2012, MNRAS, 423, 1544
\bibitem[]{Us} Scannapieco et al. 2012, MNRAS, 423, 1726
\bibitem[]{Ud1} Scannapieco, C., Tissera, P., White, S., Springel, V., 2006, MNRAS, 371, 1125
\bibitem[]{Ud12}Shen, J. \& Sellwood, J. A. 2006, MNRAS, 370, 2
\bibitem[]{Uasdf} Silk, J., Mamon, G., 2012, RAA, 12, 917
\bibitem[]{Uasf} Somerville, R., Hopkins, P., Cox, T., Robertson, B., Hernquist, L., 2008, MNRAS, 391, 481
\bibitem[]{s1} Springel V., 2005, MNRAS, 364, 1105
\bibitem[]{h1} Springel, V., et al., 2005, Nature, 435, 629
\bibitem[]{21} Springel, V., Wang, J., Vogelsberger, M., Ludlow, A., Jenkins, A., Helmi, A., Navarro, J. F., Frenk, C. S., White, S. D. M., 2008, MNRAS, 391, 1685
\bibitem[]{U1} Springel, V., 2010, MNRAS, 41, 791
\bibitem[]{U2} Stewart, K., Brooks, A., Bullock, J., Maller, A., Diemand, J., Wadsley, J., Moustakas, L., 2013, ApJ, 769, 74
\bibitem[]{r2} Tasker, E. J. 2011, ApJ, 730, 11
\bibitem[]{asf1} Tecce, T., Cora, S., Tissera, P., Abadi, M., Lagos, C., 2010, MNRAS, 408, 2008
\bibitem[]{gsf1} Vitvitska, M., Klypin, A., Kravtsov, A., Wechsler, R., Primack, J., Bullock, J., 2002, ApJ, 581, 799
\bibitem[]{UDS} Williams, R. J., Quadri, R. F., Franx, M., van Dokkum, P., \& Labb\'e, I. 2009, ApJ, 691, 1879
\bibitem[]{U1} Wuyts, S., et al., 2011, ApJ, 742, 96
\bibitem[]{york} York, D. et al., 2000, AJ, 120, 1579
\end{thebibliography}
\end{document}